%% file: ms.tex
\documentclass[preprint]{elsarticle}
\usepackage{lineno,hyperref}
\usepackage[inline]{enumitem}
\usepackage[table,xcdraw]{xcolor}
\usepackage{longtable}
\usepackage{booktabs}
\usepackage{lscape}
\usepackage{multirow}
\usepackage{verbatim}
\usepackage{todonotes}
\usepackage[gen]{eurosym}

\presetkeys%
    {todonotes}%
    {inline,backgroundcolor=yellow}{}
    
\journal{CHBR}

\newcommand\blue[1]{\textcolor{blue}{{#1}}}

\biboptions{authoryear}

\begin{document}

\begin{frontmatter}

\title{Digital Nudging with Recommender Systems: Survey and Future Directions}

\author{Mathias Jesse and Dietmar Jannach}
\address{University of Klagenfurt, Austria\\email: mathias.jesse@aau.at, dietmar.jannach@aau.at}


\begin{abstract}

Recommender systems are nowadays a pervasive part of our online user experience, where they either serve as information filters or provide us with suggestions for additionally relevant content. These systems thereby influence which information is easily accessible to us and thus affect our decision-making processes though the automated selection and ranking of the presented content. Automated recommendations can therefore be seen as digital nudges, because they determine different aspects of the choice architecture for users.
In this work, we examine the relationship between digital nudging and recommender systems, topics that so far were mostly investigated in isolation. Through a systematic literature search, we first identified 87 nudging mechanisms, which we categorize in a novel taxonomy. A subsequent analysis then shows that only a small part of these nudging mechanisms was previously investigated in the context of recommender systems. This indicates that there is a huge potential to develop future recommender systems that leverage the power of digital nudging in order to influence the decision-making of users. In this work, we therefore outline potential ways of integrating nudging mechanisms into recommender systems.

\end{abstract}

\begin{keyword}
Digital Nudging; Recommender Systems; Survey; Decision Making
\end{keyword}

\end{frontmatter}

\section{Introduction}
Automated recommendations are a common feature in today's online world. In online shops, we constantly receive recommendations regarding additional shopping items, media platforms like Spotify suggest us new tracks or artists, and even the content we see on the news feed of social media sites is tailored to our preferences in a personalized way. Such recommendations are usually designed to help users find relevant items of interest and to avoid situations of choice overload for users. At the same time, these systems support organizational goals of the providers, such as increased sales or user engagement. The broad adoption and success of recommendation technology in practice provides ample evidence that modern recommender systems are effective and can have a significant impact on the choices and decisions that consumers make. As a result, these systems can create substantial business value \citep{LeeHosanagar2018,jannachjugovactmis2019,Gomez-Uribe:2015:NRS:2869770.2843948,DBLP:journals/isr/AdomaviciusBCZ18}.

Viewed from the individual decision-making perspective, recommendations can be seen as what is called a \emph{nudge} in behavioral economics \citep{thaler2008nudge}. The concept of nudging refers to tools that help people make better decisions \emph{``without forcing certain outcomes upon anyone''} \citep{ChoiceArchitecture2014}. Originally, nudges were investigated in \emph{offline} decision scenarios, targeting mostly at decisions related to personal health or wealth. One underlying idea of nudges is that there are certain psychological phenomena that have an influence on how people make decisions. Knowing such phenomena, a \emph{choice architect} can then construct a choice environment that subtly guides people to what is considered a decision that is beneficial for them. A typical example in the literature refers to the arrangement of food in a school cafeteria. In this scenario, the goal might be to find an order that leads to the effect that students eat more healthy food.

Soon after nudging became popular, different authors used the term \emph{digital nudging} \citep{DBLP:journals/bise/WeinmannSB16,DBLP:conf/wi/MirschLJ17,DBLP:journals/cacm/SchneiderWB18}, when they transferred the concept to the online world. In a common definition, digital nudging is related to user interface (UI) elements of software applications that affect choices of the users \citep{DBLP:journals/bise/WeinmannSB16}. Again, the usual goal is to help users make decisions that are better for them, i.e., guide them to a predefined, desired option without restricting the choice space.

A recent review of digital nudging techniques in Human Computer Interaction (HCI) research can be found in \citep{23waystonudge}.
Interestingly, in this work, recommender systems were not considered as means to nudge users. 
This is to some extent surprising because a typical goal of providing recommendations is to help users making better decisions in the first place. Moreover, the presentation of selected recommendations, e.g., in the form of an ordered list, can be seen as a nudge, as it---at least \emph{implicitly}---implements different mechanisms that were identified in the literature. For example, almost all recommender systems in practice only present a specific subset of the available choices, which can be interpreted as a \emph{hiding} nudge according to the categorization of \citep{23waystonudge}. Likewise, the particular ranking of the choices corresponds to a \emph{positioning} nudge. When considering alternative categorizations, the provision of recommendations could be seen as a nudge that is based on increasing the \emph{salience} of certain options \citep{BeyondNudges202} or as an approach that increases the \emph{ease and convenience} of choosing by making some options more visible and accessible \citep{NudgingAVeryShortGuid42014}.

However, there are also opportunities to combine personalized recommendations with ``explicit'' digital nudges, an area which is largely unexplored so far in the research literature. The use of such nudges can be particularly helpful when there is more than one goal that one would like to achieve with the recommendations. Remember that the main goal of most recommender systems is to point users to choices (items) they will \emph{probably like}. However, an additional goal might be to point users to items that the system thinks the users \emph{should explore}.

For example, a recommender system for food recipes might have learned that a user likes certain types of unhealthy food. In such a situation, a ``nudging-enhanced'' recommender might first of all truthfully recommend items that are similar to those that the user liked in the past. At the same time, however, it might make use of a visible nudge to highlight individual options in the recommendation list that are healthier than the top-ranked item(s), which are probably unhealthy. Similar considerations can be made in other domains, e.g., in music recommendation. Here, a recommender might predominantly recommend tracks, genres, and artists the user liked in the past. At the same time, it might however highlight those elements that supposedly help the users to broaden their musical interests.\footnote{Today, some e-commerce sites like Amazon use visual highlighting of individual items in their recommendation lists, e.g., to mark items as being currently discounted or being ``Amazon's Choice''. In such a context, it is however not always clear if the visual highlighting represents a nudge according to the original definition, i.e., if the highlighted option is the most beneficial one for the user.}

Generally, these relationships between recommender systems and digital nudging have not been in depth explored so far. With this work, we aim to close this research gap. The contributions and the structure of this paper are as follows.
\begin{itemize}
\item We conduct a systematic review to collect a set of nudging \emph{mechanisms} (i.e., practical ways how to nudge) that were previously identified in the research literature.
\item We organize the identified 87 nudging mechanisms into a novel taxonomy, and we describe the underlying psychological phenomena that are related to the nudging mechanisms.
\item We provide a survey of research works in which nudging mechanisms were implemented in the context of recommender systems.
\item We finally highlight future perspectives of combining digital nudging approaches with recommendation systems.
\end{itemize}

The paper is organized as follows. Next, in Section \ref{sec:methodology}, we introduce the terminology used in the paper and provide details about the methodology used in the literature review. Afterwards, in Section \ref{sec:taxonomy}, we present our extended taxonomy of nudging mechanisms; the catalog of individual mechanisms is listed in Section \ref{sec:catalog}. The underlying psychological phenomena of the mechanisms are briefly discussed in Section \ref{sec:psychological-phenomena}. In Section \ref{sec:nduging-with-recommender-systems}, we finally review which mechanisms are typically present in recommender systems, and we then discuss future opportunities to add nudges to such systems.

\section{Concepts, Terminology, and Methodology}
\label{sec:methodology}
\subsection{Digital Nudging}
The concept of \emph{nudging} as a mechanism of behavioral economics was popularized by a book by Thaler and Sunstein from 2008 \citep{thaler2008nudge}. It is based on two principles, \emph{choice architecture} and \emph{libertarian paternalism}. A choice architecture is ``\emph{the environment in which people make decisions}'' \citep{ChoiceArchitecture2014}. The characteristics (or: features) of such an architecture that have an influence on which decisions people make are called \emph{nudges}. In a software application, setting a default option for a choice would be an example of a nudge. It is important to note here that, according to Thaler und Sunstein, nudges must not force people to make a certain choice. Setting default options is therefore admissible, because users are free to make a different choice. The concept of libertarian paternalism finally expresses that a nudge should be aimed at helping people make better decisions than they probably would if the nudge would not be there. Nudging, by this definition, is therefore always done to the benefit of people.

\emph{Digital nudging} is defined by Weinmann et al.~as ``\emph{the use of user-interface design elements to guide people's behavior in digital choice environments''} \citep{DBLP:journals/cacm/SchneiderWB18}. Mirsch at al.~adopt this definition and furthermore consider digital nudges as ``\emph{relatively minor changes to decision environments}''. In the online world, this decision environment is commonly represented by the user interface of an application.\cite{DBLP:conf/ecis/MeskeP17} based their definition also on that of Weinmann et al., but extend it to consider also the aspect of free decision-making. They furthermore refrain from limiting nudging to changes in the UI, but also consider choices regarding the form and content of information as potential nudges. They define digital nudging as a ``\emph{subtle form of using design, information and interaction elements to guide user behavior in digital environments, without restricting the individual’s freedom of choice.}''

\subsection{Ethical Considerations and Persuasion}
An important observation by Weinmann et al.~is that, in practice, digital nudges are not always used for the benefit of the users. They give the common example of airline websites that use nudging to sell non-essential options, which can be seen as a form of manipulation.\footnote{In a strict interpretation of Thaler and Sunstein's definition such mechanisms would therefore not be called nudges.} Similarly, Schneider et al.~observe that nudging is frequently used by marketers to sell more products and give everyday examples \citep{DBLP:journals/cacm/SchneiderWB18}. The use of digital nudging therefore soon leads to ethical questions. A general problem here is also that there is no entirely neutral way of presenting options in a user interface \citep{DBLP:journals/cacm/SchneiderWB18}, because designers always have to choose a certain visual representation that results in a certain order of the options.

Comparable ethical questions also arise in the context of recommender systems. In the best case, recommender systems create value both for consumers and providers, and eventually also for additional stakeholders \citep{JannachAdomavicius2016purpose,AbdoallahpouriMSRSUMUAI2020}. Academic research mostly focuses on the consumer perspective, e.g., by trying to develop algorithms and systems that help users find the most relevant items for them. One underlying assumption is that this, at least in the long term, is also beneficial for the recommendation provider, e.g., because the consumers develop trust over time. In reality, however, recommendation providers may consider not only the consumer value, but also try to maximize the organizational value, e.g., by building price and profit aware systems that balance the potentially competing goals \citep{JannachAdomaviciusVAMS2017}. Furthermore, some recommender systems might entirely filter out some options for users in a personalized way, e.g., because they are not considered relevant for them. In such situations, the principle of \emph{freedom of choice} in Meske and Potthoff's definition would be violated.

In the context of our present work, which is on the relation of digital nudging and recommender systems, we however do not primarily focus on ethical questions of digital nudging. In general, choice architects are advised to consider the implications of potentially unethical nudges \citep{DBLP:journals/cacm/SchneiderWB18,Sunstein2015NudgingAC}. According to Meske and Potthoff, the discussion regarding ethical considerations however has not achieved a consensus yet \citep{DBLP:conf/ecis/MeskeP17}.

The consideration of ethical questions is sometimes considered as a difference between nudging and the concept of \emph{persuasion}.
In \cite{Recommendationswithanudge2019}, they contrast nudging with other types of behavioral interventions, including persuasion. When considering recommendations as nudges, Karlsen and Andersen consider the removal of options as an aspect that is more common (and allowed) in persuasive approaches. However, note that according to a common definition of persuasive technology \citep{FoggPresuasion}, the use of coercion or deception is not acceptable.

A discussion of commonalities and differences between nudging and persuasion that goes beyond ethical aspects can be found in \cite{DBLP:conf/ecis/MeskeP17}. In the research literature, recommender systems are often considered as persuasive technology, see e.g., \cite{YooPersuasiveRS2013}. In the context of our work, we however recognize that it can be difficult to always draw a clear line between nudging and persuasion. In the end, while the concepts were developed in different disciplines, they share the same goal and aim at influencing the behavior and decisions of people \citep{DBLP:conf/ecis/MeskeP17}. In our present work, we therefore do not adopt a dogmatic approach, but accept that the concepts of nudging and persuasion are overlapping. We therefore may also include nudges in our survey that some might call ``persuasive cues'' elsewhere \citep{YooPersuasiveRS2013}.

\subsection{Methodology -- Literature Search}
To identify relevant research literature, we first conducted a systematic search on major electronic libraries. The aim was to find papers that were concerned with the application of nudges in the field of recommender systems. Based on the initial findings, citation snowballing was conducted to find relevant and related works. In order to ensure that we issued queries of the form (``Nudge'' OR "Nudging'' OR ``Digital Nudge'' OR "Digital Nudging'') AND (``Recommender'' OR ``Recommender Systems'') in the following libraries: Web of Science, SpringerLink, IEEE Xplore, Directory of Open Access Journals, Emerald Insight, ERIC MEDLINE/PubMed, Sage Journals, INFORMS Journals. We then manually scanned the titles and abstracts of the returned works to determine if they were relevant for our survey. These papers were deemed relevant when (a) they were published after the year 2008 \footnote{The term ``nudge'' was popularized in the year 2008.}, (b) described a nudge and (c) linked those to recommendations in some way. With the help of snowballing, we subsequently identified additional works on digital nudging which were not necessarily focusing on recommender systems.
This entire procedure resulted in 111 papers. 
In the end, after removing duplicates and checking for relevancy, we were left with 63 papers that we further considered in our survey.

\section{Taxonomy and Catalog of Nudging Mechanisms}
\label{sec:taxonomy}
In this section, we present the taxonomy which we used to categorize the different nudging mechanisms that we found in the literature. The developed taxonomy builds upon and refines previous categorizations found, e.g., in \cite{23waystonudge,Muenscher2016,HUMMEL201947,Mindspace2010}. We will discuss the rationale for our adaptations below.

In this context, note that the literature that proposes such taxonomies is not always consistent in terms of terminology. Some authors for example use the term \emph{choice architecture tool} \citep{ChoiceArchitecture2014,BeyondNudges202}, \emph{choice architecture technique}, or \emph{intervention technique} \citep{Muenscher2016} when they refer to different ways of nudging, e.g., by setting a default. Others use the terms \emph{nudging mechanism} \citep{23waystonudge}, \emph{nudging technique} \citep{Djurica2017TheEO}, \emph{nudge principle} \citep{DBLP:journals/cacm/SchneiderWB18}, \emph{nudging element} \citep{DBLP:conf/ecis/MeskeP17} or simply \emph{(digital) nudge} \citep{Schaer2019} when referring to more or less the same thing. In an earlier work \citep{Mindspace2010} the term \emph{influences on behavior} is used.  Furthermore, note that in existing research works we find categorizations that mix nudging mechanisms like setting a default with psychological phenomena like the anchoring or the framing effect, e.g., \citep{DBLP:conf/ecis/MeskeP17}.

In our work, we will consistently use the term \emph{nudging mechanism} in the remainder of the paper, thereby adopting the terminology of \cite{23waystonudge} and others. In addition, we will differentiate between the mechanisms and the underlying psychological effect. We provide a novel mapping of nudging mechanisms and psychological phenomena in Section \ref{sec:psychological-phenomena}.

\subsection{Taxonomy of Nudging Mechanisms}
\label{sec:taxonomy-of-nudges}

\begin{figure}[h!]
	\centering
	\includegraphics[width=1\linewidth]{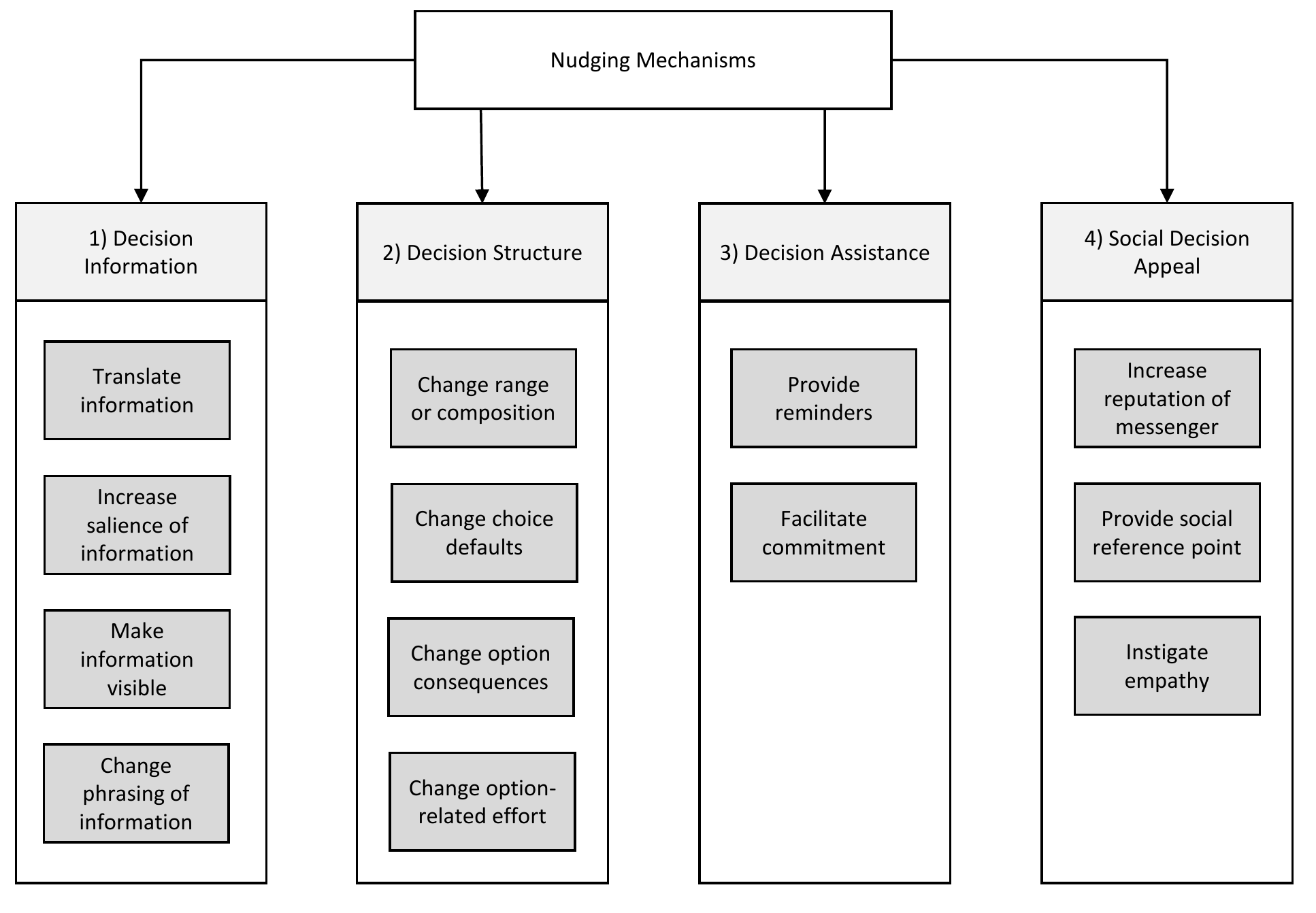}
	\caption{Taxonomy of Nudging Mechanisms}
	\label{fig:taxonomy-high-level}
\end{figure}

Figure \ref{fig:taxonomy-high-level} shows the different categories of our taxonomy of nudging mechanisms. At the highest level, we identified four main categories of nudging mechanisms: \emph{Decision Information}, \emph{Decision Structure}, \emph{Decision Assistance}, and \emph{Social Decision Appeal}.

The first three categories were proposed previously in \cite{Muenscher2016}. Based on our catalog of nudging mechanism (see Section \ref{sec:catalog}), we added a fourth category named \emph{social decision appeal} to better reflect the characteristics of the mechanisms in our taxonomy.\footnote{The categorization in \cite{HUMMEL201947} is based on only two higher-level categories, \emph{Structuring the choice task} and \emph{Describe choice options}.}

\begin{itemize}
\item \emph{Decision Information} This category of nudging mechanisms is based on changing the information that is shown to the decision maker without changing the options themselves. Such changes can, for example, be based on increasing the salience of the provided information or on reframing the description of the options.

\item \emph{Decision Structure} By altering the structure of a decision, a choice architect can steer the behavior of decision makers. Nudging mechanisms in this category focus on the arrangement of options. Examples for this are setting defaults, changing the effort, or ordering the options.

\item \emph{Decision Assistance} As the name suggests, this category encompasses mechanisms that support decision makers in accomplishing their goals. Choice architects can, for example, do this by reminding users of preferred options, or by helping users carry out plans that they previously mentioned.

\item \emph{Social Decision Appeal} Similar to the nudging mechanisms mentioned in the category of decision information, the ones in this category are also based on changing the information that is presented to users.
What sets these two apart is that \emph{social decision appeal} focuses on the emotional and social implications of the change. The underlying assumption is that decision makers are more inclined by what other individuals did in the same situation and are influenced by their decisions.
\end{itemize}

At the second level of the taxonomy, we suggest 13 subcategories based on the analyzed papers. Nine of the categories are based on the proposal by \cite{Muenscher2016}. Again, we suggest additional categories, as this allows for a fine-grained classification. These additional subcategories are \emph{Increase Salience} and \emph{Increase Reputation of Messenger}, inspired by \cite{Mindspace2010}, \emph{Use insights from Human Behavior}, to cover mechanisms that are directly related to a single psychological phenomenon, and \emph{Instigate empathy}, inspired by \cite{23waystonudge,SunsteinCouncil2016}.


Overall, with the proposed taxonomy our goal is to avoid certain limitations of previous proposals. In particular, existing taxonomies sometimes use a comparably coarse structure, which makes it difficult to classify the variety of nudging mechanisms that we found through our literature survey in an appropriate manner. For instance, we found it challenging to classify social or emotion-based mechanisms (e.g., moral suasion) in existing taxonomies in a way that allows to consider the finer details of the different types of nudging mechanisms.


\subsection{Catalog of Nudging Mechanisms}
\label{sec:catalog}
\begin{figure}[h!t]
	\includegraphics[width=1\linewidth]{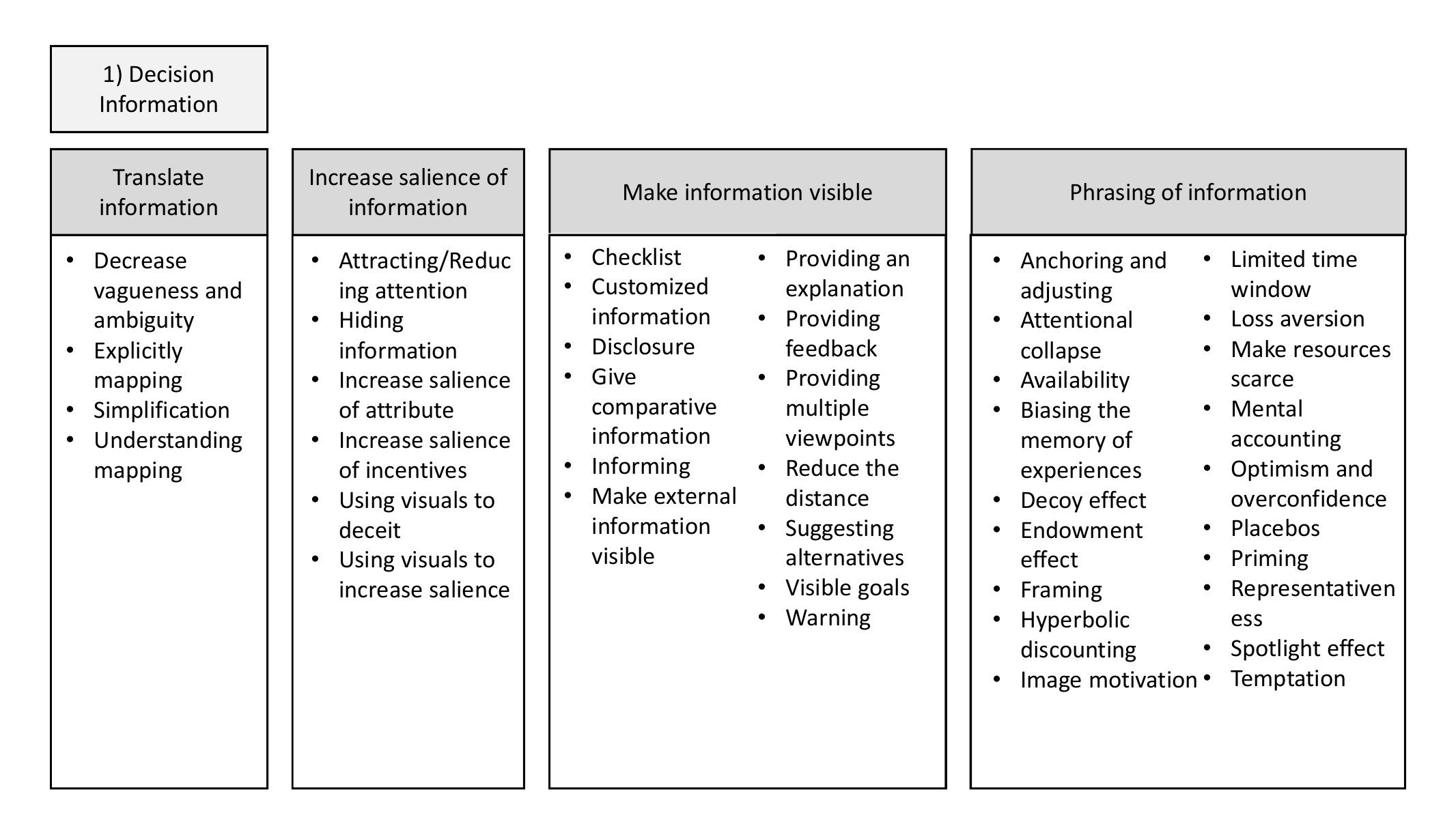}
	\caption{Overview of Nudging Mechanisms in the Category \emph{Decision Information}}
	\label{fig:taxonomy-decision-information}
\end{figure}

Next, we present our detailed catalog of nudging mechanisms. We organize the discussion according to the categories of the presented taxonomy. An overview of each category is given in Figure \ref{fig:taxonomy-decision-information} to Figure \ref{fig:taxonomy-decision-appeal}. Details for each nudging mechanisms are provided in the Appendix in Tables \ref{table:decision_information_nudges} to Table \ref{table:decision-appeal-nudges}.

\begin{itemize}
  \item \emph{Decision Information}: About half of the identified nudging mechanisms fall in this category, which is based in providing or emphasizing certain types of information to guide the decision maker.
  \item \emph{Decision Structure}: In this category, we found 23 nudging mechanisms in four subcategories.
  \item \emph{Decision Assistance}: We categorized eight nudging mechanisms under two sub-categories of \emph{Decision Assistance}.
  \item \emph{Social Decision Appeal}: Finally, there are 11 nudging mechanisms in the new category labeled \emph{Social Decision Appeal}, grouped in three sub-categories.
\end{itemize}

\begin{figure}[h!t]
	\includegraphics[width=0.9\linewidth]{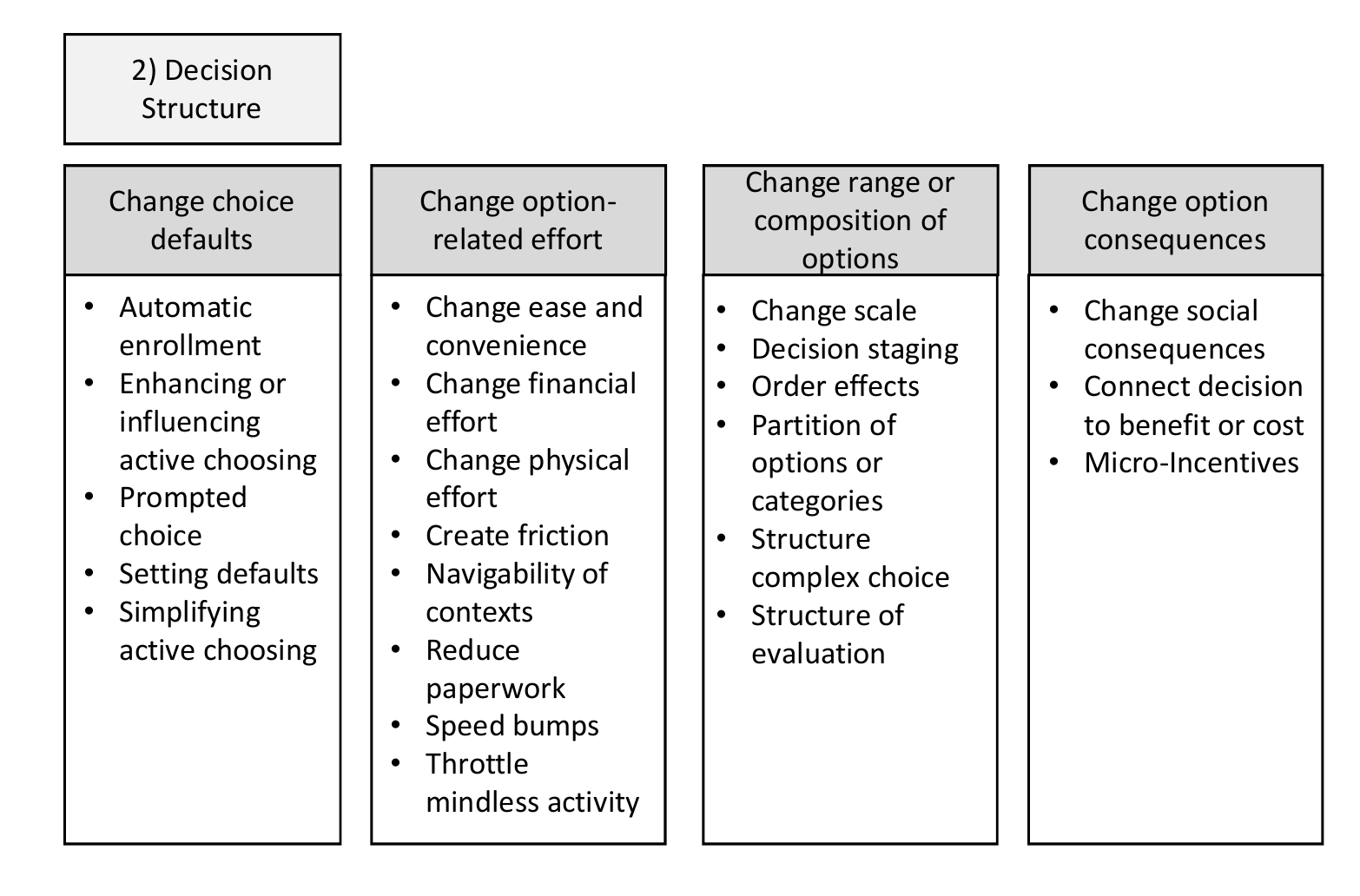}
	\caption{Overview of Nudging Mechanisms in the Category \emph{Decision Structure}}
	\label{fig:taxonomy-decision-structure}
\end{figure}

\begin{figure}[h!t]
	\includegraphics[width=.7\linewidth]{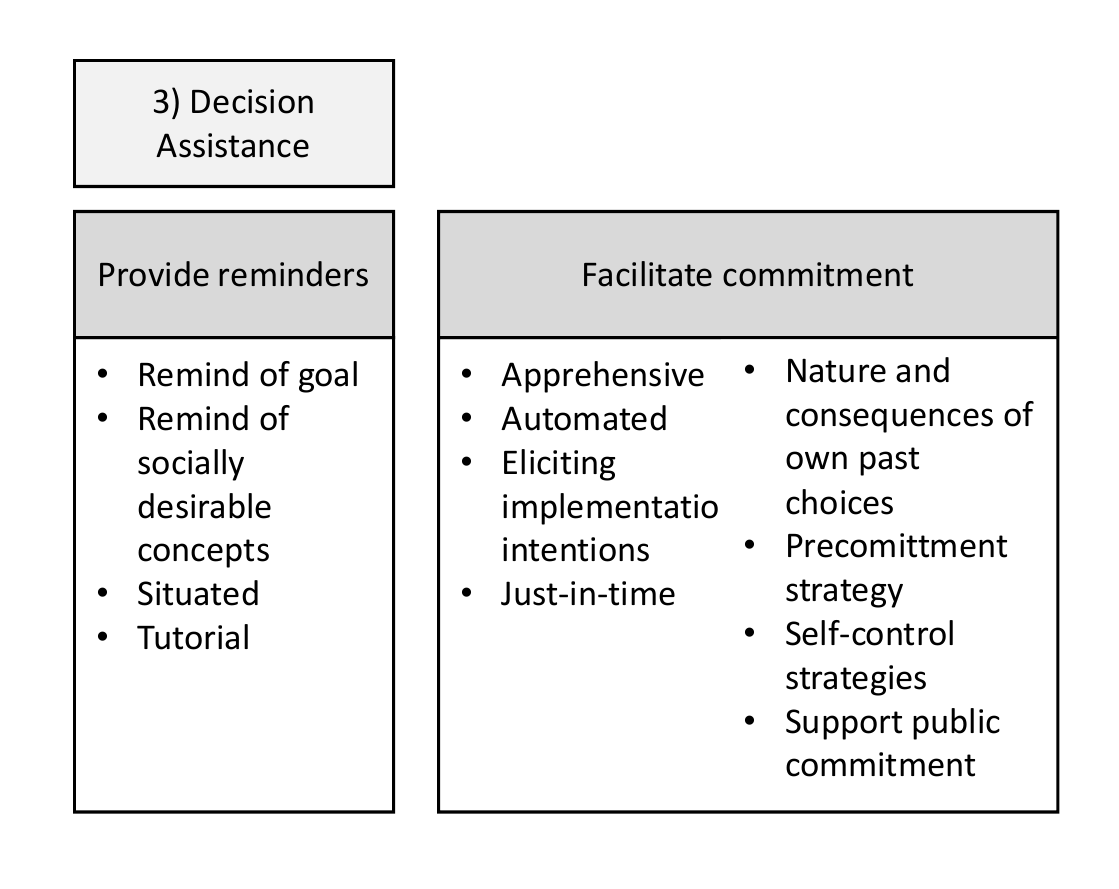}
	\caption{Overview of Nudging Mechanisms in the Category \emph{Decision Assistance}}
	\label{fig:taxonomy-decision-assistance}
\end{figure}

\begin{figure}[h!t]
	\includegraphics[width=.7\linewidth]{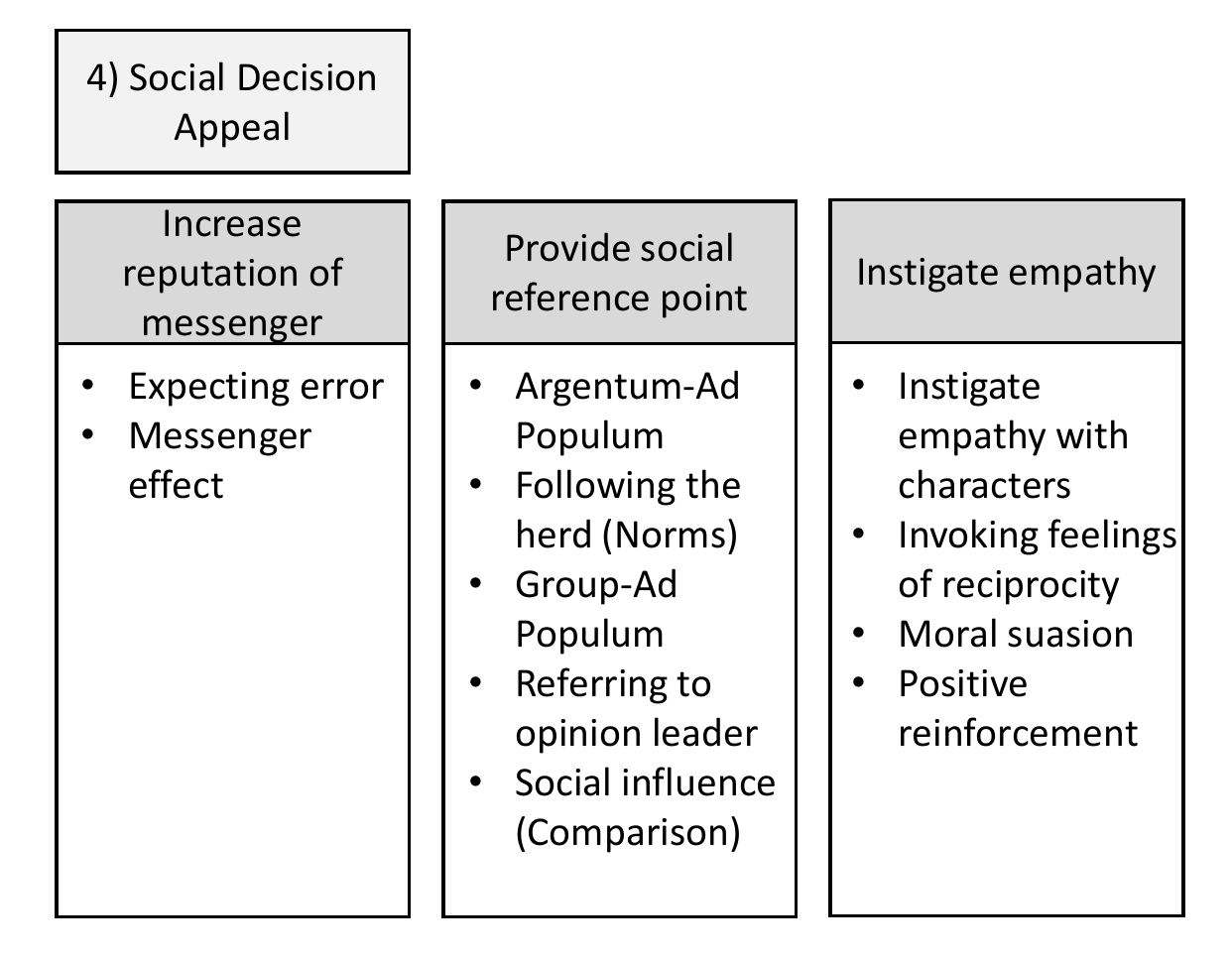}
	\caption{Overview of Nudging Mechanisms in the Category \emph{Social Decision Appeal}}
	\label{fig:taxonomy-decision-appeal}
\end{figure}

\section{Underlying Psychological Phenomena}
\label{sec:psychological-phenomena}
The principle of nudging, as proposed by \cite{thaler2008nudge}, is based on the existence of a number of psychological phenomena that can have an effect on how we make decisions. In the original work on nudging, a number of decision heuristics and biases are provided as examples, including the anchoring effect or the availability or similarity heuristic, as proposed in \cite{Tversky1124}.
Our literature review revealed that researchers refer to a variety of such underlying phenomena when characterizing the principles of nudging. In our literature review, we identified 58 psychological phenomena that were at least mentioned in the papers. We list them in Table \ref{tab:psychological-phenomena}.

\input{psychological_phenomena}

However, when individual nudging mechanisms are described in the literature, it is often not stated explicitly upon which theoretical considerations these mechanisms are assumed to rely on. 
Table \ref{tab:nudges-with-phenomena} shows those cases where the underlying phenomena were explicitly mentioned in the literature.
As it can be seen from the table, less than one third (22 of 87) nudging mechanisms in the literature are explicitly based on underlying decision-related phenomena. Our review therefore points to an open research gap, as in many cases nudging mechanisms may rather be defined based on researcher experience than on underlying psychological theories or known phenomena in the context of human decision-making. However, to make nudges as effective as possible, it is important to understand what ``makes them work'' based on a solid understanding of the underlying theories.

\input{nudges_and_effects}

As a side note, when reviewing the list of psychological effects in the literature, it became evident that these effects are sometimes used interchangeably with a corresponding nudging mechanism. Examples are mechanisms from the literature like \emph{anchoring and adjustment} or \emph{framing}, which directly refer to underlying psychological phenomena. As a result, we face the problem that sometimes nudging mechanisms are proposed without strong backing theory or empirical evidence. Furthermore, the used terminology can also lead to confusion.


In order to understand how nudging mechanisms work in recommender systems, it is necessary to understand their underlying effects. Each mechanism can be based on one or more psychological effects. The implementation of mechanisms can vary and thus create a unique set of interactions between their effects.
Furthermore, demographic characteristics can alter the perception. For example, the work of \cite{Esposito.2017} found that the effects of age are directly related to nudges. In their example, this meant that older participants were less likely to being affected by nudges.

To assess these inter-relationships between underlying psychological phenomena and the specific implementation of a nudging mechanism, the system needs to be designed and evaluated as a whole. 
Existing research suggests that the resulting effects cannot be directly connected to the mechanisms, as too many variables have to be observed at once \citep{Muenscher2016, HUMMEL201947, Dolan.2012}. This can be attributed to the facts that \emph{(i)}
multiple ways exist of implementing the same nudging mechanism and \emph{(ii)} each mechanism can leverage more than one psychological underlying phenomenon. 
Overall, further research is needed to understand the relationships between nudging mechanism and underlying psychological phenomena.


%

\section{Nudging with Recommender Systems}
\label{sec:nduging-with-recommender-systems}
In this section, we investigate the relationship between typical recommender systems and digital nudging in more depth. First, in Section \ref{subsec:recommendation-as-nudging}, we argue that online recommendations can be viewed as digital nudges and why they \emph{implicitly} implement more than one of the nudging mechanisms of our catalog. Afterwards, in Section \ref{subsec:nudging-mechansims-in-recommenders}, we report cases from the literature where recommendations were \emph{explicitly} combined with digital nudging.

\subsection{Recommendation viewed as Nudging}
\label{subsec:recommendation-as-nudging}
Recommendations, as provided in the context of various online services, have proven effects on the behavior of consumers.\footnote{See \cite{jannachjugovactmis2019} for an overview on the effectiveness of recommender systems in business environments.} As such, they can be seen as a form of behavioral intervention. In the following, we consider the case of a typical recommendation system as found on Amazon.com or at Netflix, i.e., a system that filters and ranks the available options for the consumers. Such a filtering and ranking approach \emph{inherently} implements several principles of the nudging mechanisms of our catalog. We list them in Table \ref{tab:native-mechanisms}.


\input{native_mechanisms}

These inherent mechanisms manifest themselves as follows.
Typical recommender systems aim to provide recommendations which are tailored (\emph{customized information}) towards the needs of each user and therefore reduce the physical effort (\emph{change physical effort}) in terms of searching and finding a fitting item. Another aim is to simplify the process of finding fitting items (\emph{simplification}). Furthermore, they enable users to make more complex choices (\emph{structure complex choices}). In a typical recommendation process, rarely only one recommendation is provided by the system. Based on the needs of the specific user, various options are presented for them to choose from  (\emph{suggesting alternatives}). Moreover, the recommendations are presented in a list or some other visual arrangement. This puts additional emphasis on certain items and hides unwanted options (\emph{ordering effects and hiding}). Even more so, these are arranged in lists or on different pages, that alter their perception (\emph{partition of options and categories}). Items provided by the system are usually accompanied by information like a title, tags or price (\emph{informing}). One such attribute that helps users to compare their behavior to that of others, is the majority of user actions. Usually this is expressed with what has been liked or purchased by others (\emph{following the herd and social influence}).  Each attribute and incentive can further be emphasized (\emph{increase salience of attributes and incentives}) which helps increase the ease and convenience of using the recommender system (\emph{change ease and convenience}).

\subsection{Existing Studies on Nudging with Recommender Systems}
\label{subsec:nudging-mechansims-in-recommenders}

Our systematic literature review surfaced 16 papers which reported on having implemented and analyzed nudges in recommender systems. Three additional papers proposed an implementation and described how, in the future, they would measure the effects \citep{Kissmer.2018, Brown.2019, Jung.2018}. When excluding these three papers, 18 nudging mechanisms were studied in the literature, as shown in Table \ref{tab:nudges-in-recommender}.

\vspace{10pt}
\input{nudges_in_recommenders}

\newpage
Comparing the list of all identified nudging mechanisms (87) and the ones implemented in recommender systems (18) reveals that the use of 69 nudging mechanisms proposed in the literature has not been explored yet\footnote{Note that in some works the authors did not explicitly use the term nudging when they explored the effects of a psychological phenomenon in the context of recommender systems, e.g., in \cite{Teppan.2009}. We therefore cannot rule out that other works exist that investigate nudging mechanisms with recommender systems, as they were not surfaced through our literature search.}.
Only a handful of nudging mechanisms were analyzed in more than one implementation. As can be seen in Table \ref{tab:nudges-in-recommender}, the most researched mechanisms are ``Setting defaults'' and ``Using visuals to increase salience''.

Looking closer at the works listed in Table \ref{tab:nudges-in-recommender}, we can make the following observations regarding the applied research methodology.

\begin{itemize}
  \item \emph{Application Domains:} The papers analyzed in our study focused on nine different application domains. Among these domains, route planning, e-commerce, internet service, and social media were found more than
once with route planning and e-commerce being the most popular ones.
  \item \emph{Types of Experiments:} Various experimental settings were explored to study the nudging mechanisms, including surveys, interviews, laboratory studies, or field tests. The latter two, laboratory studies and field tests, were most commonly used, sometimes paired with a pre-test before the study or a survey afterwards.
  \item \emph{Study sizes:} The study sizes in the analyzed papers range from eight \citep{Verma.2018} to around 20,933 users \citep{Gena.2019}, the latter being a clear outlier. Without two outliers, the mean number of participants is at 223 and the median is 100.
  \item \emph{Study Duration:} Only in a few papers the overall duration of the studies is documented \citep{Bothos.2015, E.Bothos.2016, Gena.2019, Forwood.2015, Esposito.2017}. In these works, the duration was either one month or two months.
  \item \emph{Dependent Variables:}  Eleven studies measured if the implemented mechanisms had any \emph{impact on the behavior of users}. The remaining papers stated that nudging mechanisms were part in their setup, but not investigate them systematically. One such paper focused on the usability of the recommender system \citep{Karvonen.2010}. Others did not conduct a user study but focused on providing conceptual insights. Such works for example analyzed a mathematical model \citep{Guers.2013, Cheng.2016} or proposed a conceptual model \citep{Bauer.2017,Knijnenburg.2014}. 
  \item \emph{Effectiveness:} The majority of the papers that included user studies, 8 out of 11, report that at least one nudging mechanism used in their implementation had an impact on the behavior of users. 
      The remaining three works could not report significant effects of nudging \citep{E.Bothos.2016,Forwood.2015, Lee.2011}. The observed effects or non-effects of the tested nudging mechanisms are summarized in Table \ref{tab:overview-papers}.
 \end{itemize}

\input{overview_papers}
\normalsize

\paragraph{Discussion} Our analysis indicates that the effects of nudging mechanisms were explored in a variety of domains, underlining that their broad applicability in practice. Regarding the study sizes, the numbers are generally encouraging. A few studies are however based on very small samples, which makes it difficult to judge how reliable and generalizable the obtained results are. Study sizes are often mentioned as a limitation to the conducted study \citep{E.Bothos.2016, Kammerer.2014,Verma.2018}.


Regarding the durations of these studies, only two papers \citep{Bothos.2015, E.Bothos.2016}, both authored by the same researchers, investigated long-term effects of nudges. All other research focused on the immediate effects of the nudging mechanisms. As a result, it remains difficult to say what the long-term effects of nudges in recommender systems really look like. As a consequence, more research is needed that concentrates on observing the changes in behavior in a longer time frame.

With regard to the effectiveness of nudging mechanisms in the context of recommender systems, the results are also encouraging, as in the majority of the studies a significant impact on user behavior could be observed.
The non-success of some mechanisms however also indicates that their effectiveness can depend on specific circumstantial situations, the application domain, or the chosen way of implementing the nudge. The investigation of the effectiveness in additional domains was for example mentioned as a direction for future work in \cite{Kammerer.2014, Forwood.2015}.

Finally, given the insights from \cite{KoecherEtAl2018} about multi-attribute anchoring effects, it might be advisable to not only measure if a nudging mechanisms led to the selection of one particular item, i.e., the nudged one. The study in \cite{KoecherEtAl2018} showed that highlighting one of several possible options as a recommendation did \emph{not} lead to the effect that this option was chosen more often. However, the recommendation had a significant effect on the choices of the users. In their studies, the recommendation of a particularly large backpack for example led to the effect that users on average selected a larger backpack from the set of available choices. Such indirect and inspirational effects of recommendations were also observed in \cite{KamehkhoschBonninJannach19} for the music domain.

\subsection{Nudging Mechanisms in Recommender Systems -- A Research Agenda}

Our study revealed that 69 nudging mechanisms were not explored yet in the context of recommender systems. We therefore analyzed which of these mechanisms could be embedded or combined with a typical recommender system in an intuitive way. In the following, we sketch a few examples of such nudging opportunities. We in most cases assume application scenarios where the original ordering of the recommendations, as returned by an underlying algorithm, is maintained when the nudging mechanisms are applied. A corresponding strategy is applied, e.g., by Amazon, where certain items in the list have additional annotations such as ``Topseller'' or ``Amazon's Choice''. While such mechanisms are already used in practice, little is known about their effectiveness in the academic literature.

\begin{itemize}
  \item \emph{Visuals to increase salience:} The provision and the quality of images for a certain option (item), can significantly influence the user's choices, as demonstrated in recent \emph{visually-aware} recommender systems \citep{he2016vbpr}. Therefore, using certain imagery can help to look one option more attractive than another one.
  \item \emph{Defaults:} Recommendations are usually ordered according to the assumed utility or relevance for the users. Nonetheless, a pre-selection of a certain option, possibly paired with additional information, seems promising to explore, in particular given the known effectiveness of the \emph{Default} nudge.
  \item \emph{Warnings:} In some domains, e.g., food recommendation, some choices are less healthy than others. Warnings that are, e.g., expressed whenever a user makes an unhealthy selection, might help to nudge users towards a healthier eating behavior.
  \item \emph{Micro-Incentives:} This mechanism consists of providing an additional (financial) incentive for certain options, e.g., reduced delivery costs. It is important, however, that the incentive does not change the rational perception of the product itself.
  \item \emph{Just-in-time, situated suggestions:} Commonly, recommendations are displayed when an online user accesses a certain service. There is, however, substantial explored potential in the use of \emph{proactive} recommendation as well \citep{DBLP:journals/jiis/RookSZ20}. Such proactive recommendations, e.g., in the form of push notifications, can be seen as nudging mechanisms by themselves. However, more research is required to understand questions related to the timing of these nudges, which can be central for their effectiveness. Factors that influence the timing can, e.g.,  include questions of the user's situational context or their current \emph{decision stage}.
  \item \emph{Social influence:} As shown by \citep{Huhn.2018}, there is a direct correlation between messages coming from subject matter experts or celebrities and the acceptance of this information. Thus, these experts can influence the attitude of other users in their intention to buy products. Consequently, this could increase sales in recommender systems and needs to be further considered in research.
  \item \emph{Instigate empathy and messenger effects:} Finally, in some situations the goal of applying nudging in recommender systems might not be to promote a particular choice, but to increase the effectiveness of a recommender as a nudging system as a whole. In these cases, the use of impersonated agents (avatars), as done previously, e.g., in \citep{Jannach2004}, could be one viable option to increase the effectiveness of the recommender.
\end{itemize}

In addition to this selection of yet unexplored nudges, we see further potential for future work to validate those nudges that were already explored previously, as discussed in Section \ref{subsec:nudging-mechansims-in-recommenders} and to explore alternative implementations, also in different application settings.
Moreover, there are some behavioral intervention mechanisms that were investigated in related fields, but not in the context of nudging.  For example, in the work of \cite{Johnson.2012}, the system created a feeling of urgency or scarcity, which puts users in a situation in which they have to act quickly and make decisions that potentially overemphasize short term rewards. Such approaches resemble the ideas of the nudging mechanisms  \emph{limited time window, making resources scarce and temptation}. Such ideas are broadly used as \emph{persuasive} tools in marketing and sales.  In the future, these mechanisms could however be further explored in a positive way as well as nudges in the context of recommender systems.


On a more general level, note that providing suitable information (e.g., mapping, tutorials or simplification) can make choices easier for users in general. Furthermore, other nudging mechanisms (e.g., messenger effect or expecting error) might also help to increase the users' trust in the recommender system, thus potentially leading to a higher adoption rate of the provided recommendations and higher decision satisfaction.

There is also untapped potential in the use of \emph{personalizing} the nudging mechanisms, both their choice and the implementation, to individual users and their contextual situation. The implementation of such ``smart nudges'' however requires that additional information about the user's current situation is available to create the highest added value for the users \citep{Recommendationswithanudge2019}.



Finally, regarding the analyzed effects, it is worth noting that no paper investigated if the participants felt manipulated or coerced by the proposed nudge. Following the original definition of nudging, it is however advisable to always check for this aspect in order to avoid any unwanted manipulation of the users.

\section{Conclusion} 
Both recommender systems and digital nudging mechanisms are proven means to influence the behavior of online users in a desired way. In this paper, we have elaborated on the relationship between these concepts, and we have emphasized that recommender systems inherently implement various nudging mechanisms. Moreover, we have identified a number of research gaps as well as future ways of incorporating additional nudging mechanisms into recommender systems to further increase their effectiveness.

\bibliography{bibfile}
\newpage
\section*{Appendix A: List of Categorized Nudging Mechanisms}
\subsection{Decision Information Nudging Mechanisms}
\input{list_decision_information_nudges}

\subsection{Decision Structure Nudging Mechanisms}
\input{list_decision_structure_nudges}
\newpage
\subsection{Decision Assistance Nudging Mechanisms}
\input{list_decision_assistance_nudges}

\subsection{Social Decision Appeal Nudging Mechanisms}
\input{list_decision_appeal_nudges}
\end{document}

%% file: psychological_phenomena.tex
\begin{center}
	\small
\begin{longtable}[c]{p{5.65cm}p{5.65cm}}
  \textbf{List of effects} & \textbf{List of effects (continued)} \\
  \hline
  \endhead
  Accent fallacy  \citep{Gena.2019} & Hyperbolic discounting \citep{DBLP:conf/wi/MirschLJ17, Theocharous.2019, Johnson.2012, Thaler.2018, Mirsch.2018, Mirsch.2018b, Lee.2011} \\
  Affect heuristic \citep{23waystonudge} & Image motivation \citep{DBLP:conf/wi/MirschLJ17} \\
  Ambiguity Aversion \citep{Theocharous.2019} & Information bias \citep{Johnson.2012} \\
  Anchoring \& Adjustment \citep{DBLP:journals/bise/WeinmannSB16, DBLP:conf/ecis/MeskeP17, DBLP:journals/cacm/SchneiderWB18, DBLP:conf/wi/MirschLJ17, Theocharous.2019, Zhang.2015,Mirsch.2018, Leitner.2015, Felfernig.2008, Mirsch.2018b, Tversky.1974}  & Intertemporal choice \citep{DBLP:conf/wi/MirschLJ17} \\
  Appeal to the majority (Argumentum Ad Populum) \citep{Gena.2019} & Loss aversion \citep{DBLP:conf/wi/MirschLJ17, Theocharous.2019, Thaler.2018, Mirsch.2018, Miller.2019, Gena.2019, Mirsch.2018b} \\
  Attentional bias \citep{DBLP:conf/wi/MirschLJ17} & Mental accounting \citep{DBLP:conf/wi/MirschLJ17, Muenscher2016, Thaler.2018} \\
  Authority bias \citep{Gena.2019} & Mere exposure effect \citep{23waystonudge} \\
  Availability heuristic \citep{DBLP:journals/bise/WeinmannSB16, DBLP:journals/cacm/SchneiderWB18, DBLP:conf/wi/MirschLJ17, Thaler.2018, 23waystonudge, Tversky.1974} & Messenger effect \citep{DBLP:conf/wi/MirschLJ17} \\
  Choice aversion \citep{Muenscher2016} & Middle-option bias \citep{DBLP:journals/cacm/SchneiderWB18} \\
  Choice supportive bias \citep{Theocharous.2019} & Optimism and overconfidence bias \citep{DBLP:conf/wi/MirschLJ17, SunsteinCouncil2016} \\
  Commitment bias \citep{23waystonudge} & Peak-end results \citep{23waystonudge} \\
  Confirmation bias \citep{23waystonudge, Theocharous.2019, Loewenstein.2017} & Placebo effect \citep{23waystonudge} \\
  Conjunction fallacy \citep{Wang.2019}& Present bias \citep{SunsteinCouncil2016, Loewenstein.2017} \\
  Contrast effect \citep{Theocharous.2019} & Primacy and recency effect \citep{DBLP:journals/cacm/SchneiderWB18, Leitner.2015, Felfernig.2008, Felfernig.2007} \\
  Correspondence bias \citep{Wang.2019} & Priming \citep{DBLP:conf/wi/MirschLJ17, Bauer.2017} \\
  Decision fatique \citep{Theocharous.2019} & Procrastination \citep{SunsteinCouncil2016} \\
  Decision inertia \citep{Jung.2018} & Reciprocity bias \citep{23waystonudge} \\
  Decoupling \citep{DBLP:conf/wi/MirschLJ17, Mirsch.2018} & Regression to the mean \citep{kahneman2011thinking} \\
  Decoy effect (Asymmetric dominance) \citep{23waystonudge, Lee.2011, DBLP:journals/cacm/SchneiderWB18, Theocharous.2019, Leitner.2015, Felfernig.2008} & Representativeness and stereotypes \citep{DBLP:journals/bise/WeinmannSB16, DBLP:journals/cacm/SchneiderWB18, DBLP:conf/wi/MirschLJ17, Tversky.1974} \\
  Default effect \citep{DBLP:journals/bise/WeinmannSB16, DBLP:conf/ecis/MeskeP17, BeyondNudges202, Johnson.2012, Loewenstein.2015} & Risk aversion \citep{Theocharous.2019} \\
  Denomination effect \citep{Muenscher2016} & Salience bias \citep{23waystonudge, Johnson.2012} \\
  Distinction bias \citep{Theocharous.2019} & Scarcity bias \citep{23waystonudge, DBLP:journals/cacm/SchneiderWB18} \\
  Diversification bias \citep{Muenscher2016} & Selective perception \citep{Theocharous.2019} \\
  Endowment effect \citep{DBLP:conf/wi/MirschLJ17, Loewenstein.2015, Thaler.2018} & Simulation heuristic \citep{Wang.2019} \\
  Framing \citep{DBLP:conf/ecis/MeskeP17, DBLP:conf/wi/MirschLJ17, Theocharous.2019, Leitner.2015, Felfernig.2008, Miller.2019} & Social desirability bias \citep{DBLP:journals/cacm/SchneiderWB18, DBLP:conf/wi/MirschLJ17, Mirsch.2018, Mirsch.2018b}\\
  Gambler's fallacy \citep{Theocharous.2019} & Spotlight effect \citep{23waystonudge, DBLP:conf/wi/MirschLJ17} \\
  Group Ad Populum \citep{Gena.2019} & Status quo bias \citep{DBLP:journals/cacm/SchneiderWB18, DBLP:conf/wi/MirschLJ17, Mirsch.2018, Mirsch.2018b, 23waystonudge, Jung.2018, Lee.2011}\\
  Halo effect \citep{Gena.2019} & Sunk-cost fallacy \citep{DBLP:conf/wi/MirschLJ17, Thaler.2018} \\
  Herd instinct bias \citep{23waystonudge} & \\
  Hindsight bias \citep{Thaler.2018} & \\
  \hline
	\caption{Overview of underlying psychological phenomena}
\label{tab:psychological-phenomena}
\end{longtable}
\end{center} 

%% file: nudges_and_effects.tex
\begin{table}[h!t]
\small
\centering
\begin{tabular}{p{5cm}p{6.3cm}}
  \textbf{Nudging Mechanism} & \textbf{Psychological Effect} \\
  \hline
  Anchoring and adjustment & Anchoring and adjustment \citep{Jung.2018} \\
  Argentum Ad Populum & Argentum Ad Populum \citep{Gena.2019} \\
  Availability & Availability heuristic \citep{thaler2008nudge} \\
  Biasing the memory of experiences & Peak-end Rule \citep{23waystonudge} \\
  Deceptive visualizations &  Salience bias \citep{23waystonudge} \\
  Framing & Framing \citep{DBLP:conf/ecis/MeskeP17}\\
  Group Ad Populum & Group Ad Populum \citep{Gena.2019}\\
  Invoking feelings of reciprocity & Reciprocity bias \citep{23waystonudge} \\
  Limited time window & Scarcity effect \citep{Johnson.2012} \\
  Loss aversion & Loss aversion \citep{Mirsch.2018b} \\
  Make resources scarce & Scarcity effect \citep{23waystonudge}\\
  Optimism and overconfidence & Optimism and overconfidence bias \citep{SunsteinCouncil2016} \\
  Partition of options/categories & Mental accounting, diversification bias and denomination effect \citep{Muenscher2016}\\
  Placebos & Placebo effect \citep{23waystonudge} \\
  Priming & Mere exposure effect \citep{23waystonudge} \\
  Providing an explanation & Information bias \citep{Johnson.2012} \\
  Referring to opinion leader & Herd instinct bias \citep{23waystonudge} \\
  Representativeness & Representativeness and stereotypes \citep{thaler2008nudge} \\
  Increase salience & Salience bias \citep{23waystonudge}\\
  Setting Defaults (Status Quo Bias) &  Status Quo Bias, Loss Aversion, Decision inertia \citep{DBLP:journals/bise/WeinmannSB16, DBLP:conf/ecis/MeskeP17, BeyondNudges202} \\
  Social Influence (Comparison) & Herd instinct bias \citep{23waystonudge} \\
  Spotlight effect & Spotlight effect \citep{23waystonudge} \\
  Support public commitment & Commitment bias \citep{23waystonudge} \\
  \hline
\end{tabular}
\caption{Explicit statements in the literature about underlying psychological phenomena.}
\label{tab:nudges-with-phenomena}
\end{table} 

%% file: native_mechanisms.tex
\begin{table}[h!t]
	\small
	\centering
	\begin{tabular}{p{5.65cm}p{5.65cm}}
		\textbf{Inherent nudging mechanisms} & \textbf{Inherent nudging mechanisms (cntd.)} \\
		\hline
		Change ease and convenience & Informing  \\
		Change physical effort &  Ordering effects  \\
		Customized information & Partition of options and categories \\
		Following the herd & Simplification \\
		Hiding & Social influence \\
		Increase salience of attributes & Structure complex choices \\
		Increase salience of incentives & Suggesting alternatives \\
		\hline
	\end{tabular}
	\caption{Nudging Mechanisms Inherent to Recommender Systems}
	\label{tab:native-mechanisms}
\end{table} 

%% file: nudges_in_recommenders.tex
\begin{table}[h!t]
\small
\centering
\begin{tabular}{p{6cm}p{5.3cm}}
  \textbf{Nudging Mechanism} & \textbf{Explored in} \\
  \hline
   Argentum-Ad Populum & \citep{Gena.2019} \\
   Decoy effect & \citep{Teppan.2009, Lee.2011}\\
   Framing & \citep{Gena.2019,Guers.2013}\\
   Group-Ad Populum & \citep{Gena.2019} \\
   Increase salience of attributes & \citep{Elsweiler.2017}\\
   Increase salience of incentives & \citep{Turland.2015} \\
   Informing & \citep{Lee.2011,Esposito.2017}\\
   Make external information visible & \citep{Cheng.2016,Guers.2013,Verma.2018}\\
   Order effects & \citep{Kammerer.2014, Turland.2015}\\
   Priming & \citep{Bauer.2017}\\
   Providing feedback & \citep{Bothos.2015} \\
   Remind of goal & \citep{E.Bothos.2016} \\
   Setting defaults & \citep{E.Bothos.2016, Bothos.2015, Bauer.2017, Lee.2011} \\
   Structure complex choice & \citep{Bothos.2015} \\
   Suggesting alternatives & \citep{Forwood.2015} \\
   Support public commitment & \citep{Brown.2019} \\
   Using visuals to increase salience & \citep{Elsweiler.2017, Gena.2019, Karvonen.2010, Esposito.2017} \\
   Warning & \citep{Esposito.2017}\\
  \hline
\end{tabular}
\caption{Studies on Nudging Mechanisms in Recommender Systems}
\label{tab:nudges-in-recommender}
\end{table} 

%% file: overview_papers.tex
	\small
	\begin{longtable}[c]{p{3cm}p{8cm}}
		\textbf{Paper} & \textbf{Nudging Goal and Finding} \\
		\hline
		\endhead
  \textbf{Positive effects} & \\
  \hline
  \cite{Bothos.2015} & \textbf{Nudging goal:} To nudge users towards following routes that are more environmentally friendly with the help of messages. \newline \textbf{Finding:} Suggestions were frequently accepted and left the users more satisfied. \\
  \cite{Elsweiler.2017} & \textbf{Nudging goal:} To make users choose healthier (low fat) recipes in contrast to other unhealthy recipes by using visual nudges. \newline \textbf{Finding:} Through the use of nudges, 62,2\% of decisions resulted in healthier recipes being chosen. \\
 \cite{Esposito.2017}& \textbf{Nudging goal:} To reduce incompatible purchases by providing additional information and warnings on items. \newline \textbf{Finding:} The basic warning nudge does not change the behavior of users, but if it is phrased in an emotional way, it worked. Furthermore, while the use of logos did not decrease the incompatible purchases but providing additional information did. \\
  \cite{Gena.2019} & \textbf{Nudging goal:} To promote specific news articles and thereby capture the interest of users. \newline \textbf{Finding:} Ad Populum and Group-Ad Populum nudges were not effective in the context of news recommendation. Negative framing, on the other hand, worked. \\
  \cite{Kammerer.2014}& \textbf{Nudging goal:} To get people to click on the first element of a recommendation. \newline \textbf{Finding:} Ordering effects in a list had an impact on the chosen websites and the time spent on them. However, using a grid layout substantially reduced these effects. \\
 \cite{Teppan.2009}& \textbf{Nudging goal:} To steer the user to select a certain option more often by using decoy options. \newline \textbf{Finding:} Decoy effects were observed and had an impact on the behavior of users. \\
  \cite{Turland.2015}& \textbf{Nudging goal:} To increase the use of more secure networks by recommending and highlighting more secure alternatives. \newline \textbf{Finding:} Nudges had a significant impact on the selection of networks. 
  Increasing the salience of options with colors had a significant impact, whereas ordering effects had not. The combination of both created the strongest behavior changes. \\
  \cite{Verma.2018}& \textbf{Nudging goal:} To make users consume more relevant information by removing unsuitable content from their Twitter feeds.  \newline \textbf{Finding:} The reduction of noise in timelines of social media had a positive impact on users. Also, the use of such a nudge was perceived to be useful. \\
  \hline
  \newpage
  \textbf{No effects} & \\ \hline
  \cite{E.Bothos.2016}& \textbf{Nudging goal:} To nudge users towards following routes that are more environmentally friendly with the help of messages. \newline \textbf{Finding:} No changes in the users' behaviour could be detected or attributed to the presented messages. \\
  \cite{Forwood.2015}& \textbf{Nudging goal:} To decrease the energy density (calories per gram of food) of a purchase  by providing alternative food choices. \newline \textbf{Finding:} Providing food alternatives did not lead to a significant change to the overall energy density of what participants bought. \\
  \cite{Lee.2011}& \textbf{Nudging goal:} To promote healthy snack choices by recommending suitable alternatives. \newline \textbf{Finding:} The implemented decoy-based nudging mechanisms did not significantly change the snack choices of users. \\
  \hline
	\caption{Studies on the effects of nudging mechanisms on user behavior in the context of recommender systems.}
	\label{tab:overview-papers}
	\end{longtable} 

%% file: list_decision_information_nudges.tex
\begin{center}
    \small
\begin{longtable}[c]{p{5cm}p{6.3cm}}
	\textbf{Mechanism} & \textbf{Description} \\
	\hline
	\endhead
	\textbf{\textit{Translate information}} &  \\
	Decrease vagueness and ambiguity \citep{SunsteinCouncil2016} & By reducing ambiguity and making information easily understandable, it becomes clearer what should be expressed and thus catches the attention of users more easily. \\
	Explicitly   mapping \citep{Johnson.2012, Muenscher2016, 23waystonudge} & When users are presented with an option it is beneficial to explain the selection in regard of its possible outcomes (e.g., costs and benefits). \\
	Simplification \citep{A.Alslaity.2019, DBLP:conf/ecis/MeskeP17, NudgingAVeryShortGuid42014, Muenscher2016, HUMMEL201947, SunsteinCouncil2016} & Reduce the cognitive effort which is necessary to understand certain information or tasks by making it shorter or phrased more simple. \\
	Understanding mapping \citep{DBLP:journals/bise/WeinmannSB16, ChoiceArchitecture2014, SunsteinCouncil2016}& Explain information, which is difficult to understand, to concepts or information that are more familiar. Examples for this are an analogy or using visual aids. \\
	& \\
	\textbf{\textit{Increase salience of information}} &  \\
	Attracting/Reducing attention \citep{SunsteinCouncil2016} & A mechanism which tries to draw attention to certain options or information with the use of highlighting. \\
	Hiding   information \citep{23waystonudge} & Make undesirable options or information harder to see. \\
	Increase salience of attribute \citep{SunsteinCouncil2016, Johnson.2012} & The act of making an attribute more salient (e.g., weight, price or color)  \\
	Increase   salience of incentives \citep{BeyondNudges202, DBLP:journals/bise/WeinmannSB16} & Make incentives  more salient or visible, so they are more effective and prominent. \\
	Using visuals to deceit \citep{23waystonudge} & Through the use of optical illusions the perception and judgment of options is altered, so they appear more salient than they actually are. \\
	Using visuals to increase salience \citep{SunsteinCouncil2016, Gena.2019, Karvonen.2010} & The salience of information is increased with the use of visual effects (e.g., colors, pictures, signs or fonts). \\
	& \\
	\textbf{\textit{Make information visible}} &  \\
	Checklist \citep{SunsteinCouncil2016} & The use of   checklists helps users to understand their progress and what further needs to  be done for the decision to be finished. \\
	Customized   information \citep{DBLP:conf/ecis/MeskeP17, BeyondNudges202} & Tailor the provided information to the needs of users and thus reduce cognitive   overload. \\
	Disclosure \citep{SunsteinCouncil2016, NudgingAVeryShortGuid42014, HUMMEL201947} & Disclose related and relevant information to a certain option (e.g., full costs of a credit card) \\
	Give comparative information \citep{SunsteinCouncil2016} & Provide the users with information that is comparative to a point of view. This allows users to make more profound decisions. \\
	Informing \citep{DBLP:conf/ecis/MeskeP17, BeyondNudges202, Loewenstein.2017, Lee.2011} & A mechanism which aims to simply provide information to the users. \\
	Make external information visible \citep{Muenscher2016, Szaszi.2018, Cheng.2016, Guers.2013} & Provide information that a third party has created (e.g., information about hygiene in a coffee shop) \\
	Providing an explanation \citep{BeyondNudges202} & Provide additional information on the situation the users are in for it to be more   understandable. \\
	Providing   feedback \citep{DBLP:journals/bise/WeinmannSB16,ChoiceArchitecture2014,thaler2008nudge,HUMMEL201947,23waystonudge} & Users are given feedback on how they are performing (e.g., doing tasks right or making   mistakes) \\
	Providing  multiple viewpoints \citep{23waystonudge} & Provide an unbiased overview over a certain decision in order for users to assess the problem at hand. For example a smartphone can be viewed from different perspectives that focus on certain usages like entertainment, office related tasks or ease of use. \\
	Reduce the   distance \citep{23waystonudge} &  If a situation is too far in the future or uncertain to happen, users do not feel the urgency to act. Therefore, it is possible to present the users with similar situations to reduce the psychological distance (e.g., heavy rain sounds played in the background when acquiring a flood insurance) \\
	Suggesting  alternatives \citep{23waystonudge, Forwood.2015} & Provide the users with alternatives that they might not have considered at this point (e.g., cheaper camera with the same resolution). \\
	Visible goals \citep{Cheng.2016,MimmiCastmo.2018} & Allow users to easily assess their progress and performance against a goal state (e.g., finishing a registration). \\
	Warning \citep{SunsteinCouncil2016,DBLP:conf/ecis/MeskeP17,NudgingAVeryShortGuid42014, HUMMEL201947, Loewenstein.2017} & Warn the users with the help of visuals or other means that emphasize the problem at hand. \\
	& \\
	\newpage
	\textbf{\textit{Phrasing of information}} &  \\
	Anchoring and adjusting \citep{DBLP:conf/ecis/MeskeP17, SunsteinCouncil2016, thaler2008nudge, Mirsch.2018b,Schaer2019,Jung.2018} & Set an anchor in a decision for users to influence their judgement of the whole  decision. An anchor can be understood as a starting point, which can be adjusted to help users to make estimations of the latter provided information (e.g. showing the price of a product before reduction). \\
	Attentional  collapse \citep{Schaer2019} & Perception of users is affected by selective factors of their attention. This means users are not always aware of every information they are presented. \\
	Availability \citep{thaler2008nudge,Schaer2019} & Users assess the likelihood of a personal situation by how easy it is for them to remember   situations that are alike (e.g., car accidents). \\
	Biasing the memory of experiences \citep{23waystonudge} & Through changing the endings of events, the memory of said event can be altered. (e.g. increasing the speed of a progress bar towards the end) \\
	Decoy effect \citep{23waystonudge, Teppan.2009, Lee.2011} & This mechanism can be facilitated by adding an inferior alternative to a set of options to make the  other seem higher in value. \\
	Endowment   effect \citep{23waystonudge, Schaer2019} & Users are more likely to keep an item if they already own it than they are to acquire it if   they do not own it. \\
	Framing \citep{SunsteinCouncil2016, DBLP:conf/ecis/MeskeP17, Muenscher2016, Loewenstein.2017, Gena.2019, Schaer2019, Knijnenburg.2014, Guers.2013, 23waystonudge, Jung.2018} & It is possible to describe or frame a good option in a bad way and vice-versa. \\
	Hyperbolic   discounting \citep{Mirsch.2018b, Schaer2019, Lee.2011} & Decoupling the purchase from the payment lowers the purchase barrier and shortens the   decision-making process (e.g., buying products with a credit card). \\
	Image   motivation \citep{Schaer2019} & Actions that are favorable are attributed to ourselves as in comparison things that take a bad turn are considered the fault of others.  \\
	Limited time window \citep{DBLP:conf/ecis/MeskeP17, BeyondNudges202, Zhang.2015} & When an option is presented as only being available for a certain amount of time it is   perceived as more important and scarce. \\
	Loss aversion \citep{thaler2008nudge, SunsteinCouncil2016, Mirsch.2018b, Schaer2019}& Describes the phenomena of users weighting losses higher than winnings (e.g., the loss of \euro{100} is worse than winning \euro{100})  \\
	Make resources   scarce \citep{23waystonudge} & Announcing limited availability of an option increases the probability of users   committing to choosing it. \\
	Mental   accounting \citep{Schaer2019} & Users group payments in fictitious accounts (e.g., presents, vacation or stocks). This helps track incomes and expenses but also leads to misconceptions about how money is actually spent. \\
	Optimism and overconfidence \citep{thaler2008nudge, SunsteinCouncil2016, Schaer2019}& Users tend to underestimate challenges, costs and timescales but overestimate the rewards. \\
	Placebos \citep{23waystonudge} & The presentation of an element that has no effect upon the individual's condition, is able to change the behavior and well-being of a user due to its perceived effect (e.g., pills made of sugar). \\
	Priming \citep{SunsteinCouncil2016, DBLP:conf/ecis/MeskeP17, thaler2008nudge, Schaer2019, Bauer.2017, 23waystonudge} & The users are influenced by a certain stimulus (smell of pizza) and more inclined to behave more predictable on another process afterwards (baking pizza for dinner).  \\
	Representativeness \citep{Schaer2019, thaler2008nudge} & When shown a member of a group the users expect the member to have certain characteristics   without having actual information about it (e.g., stereotypes in a group of people) \\
	Spotlight   effect \citep{thaler2008nudge,Schaer2019, 23waystonudge} & Users tend to overestimate the significance of their own actions. \\
	Temptation \citep{thaler2008nudge} & Options that produce a quick reward are more appealing to users than those that do so over a long period of time. \\
	\caption{Nudging Mechanisms in the Category \emph{Decision Information}}
	\label{table:decision_information_nudges}
\end{longtable}
\end{center} 

%% file: list_decision_structure_nudges.tex
\begin{center}
	\small
	\begin{longtable}[c]{p{5cm}p{6.3cm}}
	\textbf{Mechanism} & \textbf{Description} \\
	\hline
	\endhead
	\textbf{\textit{Change choice defaults}} &  \\
	Automatic enrollment \citep{23waystonudge, SunsteinCouncil2016} & Automatically enrolling users, so they have to take action to get out of a certain condition (e.g., organ donor) \\
	Enhancing or influencing active choosing \citep{SunsteinCouncil2016} & This mechanism combines the nudge of prompting the users to actively choose and another nudge (e.g., order effects or loss aversion) to influence choice. \\
	Prompted choice \citep{Muenscher2016, SunsteinCouncil2016} & By making users actively choose an option they are more invested in the decision and have to actively think about the options. \\
	Setting   defaults \citep{DBLP:journals/bise/WeinmannSB16, DBLP:conf/ecis/MeskeP17, SunsteinCouncil2016, BeyondNudges202, Johnson.2012, Loewenstein.2015, NudgingAVeryShortGuid42014, Muenscher2016, HUMMEL201947, Thaler.2018, Loewenstein.2017, Szaszi.2018, Mirsch.2018b, Schaer2019, Bauer.2017, Knijnenburg.2014, 23waystonudge, Jung.2018, Lee.2011} & When options are preselected in a decision, users are more inclined to stay with the status-quo (e.g., installation wizard). \\
	Simplifying active choosing \citep{SunsteinCouncil2016} & By making active choosing more simple it increases the chances of users thinking about their own decision and the consequences entailed by it. \\
	& \\
	\textbf{\textit{Change option-related effort}} &  \\
	Change ease and convenience \citep{SunsteinCouncil2016, NudgingAVeryShortGuid42014} & The main focus of this mechanism lies on the visibility and accessibility of options (e.g., make low-cost options or healthy foods visible). \\
	Change   financial effort \citep{HUMMEL201947, Muenscher2016} & Choice architects can intervene with the perception of financial effort. Examples include postponing costs to the future without changing the actual final costs. \\
	Change   physical effort \citep{HUMMEL201947,Muenscher2016} & This mechanism changes the necessary physical effort for certain options in order for them to be more attractive/unattractive. As an example an apple would be positioned easier to grasp than less healthy alternatives. \\
	Create friction \citep{23waystonudge} & Attempt to minimize intrusiveness while maintaining the capacity to change user's behaviour (e.g., drop bike keys when car key is selected). \\
	Navigability of contexts \citep{SunsteinCouncil2016} & Make contexts or policies easily navigable for users (e.g., pointers and guides) \\
	Reduce   paperwork \citep{SunsteinCouncil2016}& Reduce the involved paperwork in completing a task. \\
	Speed bumps \citep{SunsteinCouncil2016} & Provide literal  or figurative speed bumps or cooling-off periods for users to slow down. \\
	Throttle mindless activity \citep{23waystonudge} & Reduce the amount of mindless actions the users can take or give them possible ways of making up their mistakes (e.g., a time buffer to reverse an   action). \\
	& \\
	\newpage
	\textbf{\textit{Change range or composition of options}} &  \\
	Change scale \citep{Johnson.2012} & By expanding the denominator, numerators appear larger and this in turn makes the differences   between alternatives seem larger (e.g., \euro{1} and \euro{2} versus 100 cents and 200 cents). \\
	Decision  staging \citep{Johnson.2012, DBLP:conf/ecis/MeskeP17} & Break down a complex decision into smaller ones that only focus on small subsets of   attributes that can be compared more easily (e.g., quality of camera, weight and price). \\
	Order effects \citep{SunsteinCouncil2016, 23waystonudge} & What users see first on a website or in a room makes a difference, as items perceived at first   are intuitively of higher importance than others seen later. \\
	Partition of   options/categories \citep{Johnson.2012, Muenscher2016} & Options or categories can be arranged with their respective alternatives to create a tactical representation (e.g., segregating healthy options into more diverse categories) \\
	Structure complex choices \citep{SunsteinCouncil2016, BeyondNudges202, DBLP:journals/bise/WeinmannSB16} & The aim of this mechanism is to make a complicated choice easier to decide by restructuring it into smaller more manageable pieces. This could be done by guiding users through a process. \\
	Structure of evaluation \citep{SunsteinCouncil2016} & The way users are presented with results can influence their behavior. Assessing a single item at a time in comparison to all at once will result in different decisions. \\
	& \\
	\textbf{\textit{Change option consequences}} &  \\
	Change social consequences \citep{Muenscher2016} & Actions taken by users are linked to social outcomes (e.g., friends know about the   decision/action) \\
	Connect   decision to benefit/cost \citep{Muenscher2016} & Connecting small benefits to behavior that should be pursued or small costs to behavior can change the occurrences of outcomes. These benefits or costs must be so small as to not change the economic incentives of each option (e.g. emotional incentive).  \\
	Micro-Incentives \citep{HUMMEL201947} & Micro-Incentives are changes to the consequences of an option that are insignificant from a   rational choice perspective (e.g. free shopping bag that would usually cost three cents)
	\label{table:decision_structure_nudges}\\
	\caption{Nudging Mechanisms in the Category \emph{Decision Structure}}
\end{longtable}
\end{center} 

%% file: list_decision_assistance_nudges.tex
\begin{center}
	\small
	\begin{longtable}[c]{p{5cm}p{6.3cm}}
		\textbf{Mechanism} & \textbf{Description} \\
		\hline
		\endhead
		\textbf{\textit{Provide reminders}} &  \\
		Remind of goal \citep{SunsteinCouncil2016, DBLP:conf/ecis/MeskeP17, NudgingAVeryShortGuid42014, Brown.2019, Muenscher2016, HUMMEL201947, Szaszi.2018, 23waystonudge} & Tell the users what the result of finishing a process is. Remind them of why they should finish it. \\
		Remind of socially desirable concepts \citep{Muenscher2016} & Influence users by stating the social expectation in a specific situation (e.g., need to vote) \\
		Situated \citep{Brown.2019} & Make suggestions to the users where the practice is applicable and needed \\
		Tutorial \citep{Brown.2019}& Show users how a certain procedure works, so they can do it on their own. This could go as far as providing helps and further instructions. \\
		& \\
		\textbf{\textit{Facilitate commitment}} &  \\
		Apprehensive \citep{Brown.2019} & Providing multiple locations where the same decision can be performed by the users (e.g., different routes lead to the same result). \\
		Automated \citep{Brown.2019} & Perform actions for the users and present the results without them having to do anything.\\
		Eliciting   implementation intentions \citep{NudgingAVeryShortGuid42014,HUMMEL201947,SunsteinCouncil2016} & Users are more likely to continue an action if someone elicits their implementation   intentions (e.g., "Do you plan to vote?"). \\
		Just-in-time \citep{Brown.2019, 23waystonudge} & Make recommendations when an action is appropriate (e.g., when the user is actively searching for a product, ask if they need assistance).\\
		Nature and consequences of own past choices \citep{NudgingAVeryShortGuid42014, SunsteinCouncil2016}& With existing   historical data it is possible to explain the users why they took a certain course of actions in the past and why they should do it again. \\
		Precomitment   strategy \citep{SunsteinCouncil2016, DBLP:conf/ecis/MeskeP17, NudgingAVeryShortGuid42014, Muenscher2016,HUMMEL201947} & Let users decide on a certain course of actions or let them define a goal to work towards. \\
		Self-control   strategies \citep{thaler2008nudge} & As the name   suggests the goal of this mechanism is to support users in fulfilling their plan without falling victim to their own weaknesses (e.g., heuristics or biases). \\
		\newpage
		Support public   commitment \citep{Brown.2019, Muenscher2016, 23waystonudge} & Publicly   committing to a behavior (working out) or to refrain from a behavior   (smoking) creates external pressure. This leads to fewer people breaking their acting against their commitments.
		\label{table:decision_assistance_nudges}\\
		\caption{Nudging Mechanisms in the Category \emph{Decision Assistance}.}
	\end{longtable}
\end{center} 

%% file: list_decision_appeal_nudges.tex
\begin{center}
	\small
	\begin{longtable}[c]{p{5cm}p{6.3cm}}
	\textbf{Mechanism} & \textbf{Description} \\
	\hline
	\endhead
	\textbf{\textit{Increase reputation of messenger}} &  \\
	Expecting error \citep{BeyondNudges202, DBLP:journals/bise/WeinmannSB16} & Users make errors and by being as forgiving of that as possible, improves the reputation of the   messenger. \\
	Messenger effect \citep{Schaer2019} &  Every piece of information is delivered by a messenger (e.g., human, computer, recommender system). This messenger creates a certain picture in the mind of users and therefore affects the perceiption of it.\\
	& \\
	\textbf{\textit{Provide social reference point}} &  \\
	Argentum-Ad Populum \citep{Gena.2019} & Accepting a certain thesis based on the mere fact that the majority of people accept it. \\
	Following the herd (Norms) \citep{SunsteinCouncil2016, Mirsch.2018b, Schaer2019, thaler2008nudge} & As users do not want to stick out, they have a tendency to do what the majority of people does. \\
	Group-Ad Populum \citep{Gena.2019} & Accepting a certain thesis based on the mere fact that a specific group of people accept it. \\
	Referring to opinion leader \citep{Muenscher2016} & Provide a highly valued or respected messenger to recommend items in order to influence the opinions and behavior of users. \\
	Social influence (Comparison) \citep{DBLP:conf/ecis/MeskeP17, NudgingAVeryShortGuid42014, Muenscher2016, HUMMEL201947}& Allow users to compare their actions to that of other users. \\
	& \\
	\textbf{\textit{Instigate empathy}} &  \\
	Instigate empathy with characters \citep{23waystonudge} & Empathy can created in multiple ways, one of which is to make use of characters. This is done by showing an avatar or character that changes its state based on the behavior of a user. \\
	Invoking feelings of reciprocity \citep{23waystonudge} & If something good happens to users they try to reproduce this feeling towards the benefactor. \\
	Moral suasion \citep{SunsteinCouncil2016} & Moral suasion is used to increase fun or create a sense of responsibility which the users might act   on. \\
	Positive reinforcement \citep{Brown.2019} & Commending users when they align with the expected behavior increases empathy and further motivates them.
	\label{table:decision-appeal-nudges}\\
	\caption{Nudging Mechanisms in the Category \emph{Social Decision Appeal}.}
\end{longtable}
\end{center} 

%% file: ms.bbl
\begin{thebibliography}{71}
\expandafter\ifx\csname natexlab\endcsname\relax\def\natexlab#1{#1}\fi
\providecommand{\url}[1]{\texttt{#1}}
\providecommand{\href}[2]{#2}
\providecommand{\path}[1]{#1}
\providecommand{\DOIprefix}{doi:}
\providecommand{\ArXivprefix}{arXiv:}
\providecommand{\URLprefix}{URL: }
\providecommand{\Pubmedprefix}{pmid:}
\providecommand{\doi}[1]{\href{http://dx.doi.org/#1}{\path{#1}}}
\providecommand{\Pubmed}[1]{\href{pmid:#1}{\path{#1}}}
\providecommand{\bibinfo}[2]{#2}
\ifx\xfnm\relax \def\xfnm[#1]{\unskip,\space#1}\fi
\bibitem[{{A. Alslaity} \& {T. Tran}(2019)}]{A.Alslaity.2019}
\bibinfo{author}{{A. Alslaity}}, \& \bibinfo{author}{{T. Tran}}
  (\bibinfo{year}{2019}).
\newblock \bibinfo{title}{Towards persuasive recommender systems}.
\newblock In {\it \bibinfo{booktitle}{2019 IEEE 2nd International Conference on
  Information and Computer Technologies (ICICT)}\/} (pp.
  \bibinfo{pages}{143--148}).
\newblock \DOIprefix\doi{10.1109/INFOCT.2019.8711416}.
\bibitem[{Abdollahpouri et~al.(2020)Abdollahpouri, Adomavicius, Burke, Guy,
  Jannach, Kamishima, Krasnodebski \& Pizzato}]{AbdoallahpouriMSRSUMUAI2020}
\bibinfo{author}{Abdollahpouri, H.}, \bibinfo{author}{Adomavicius, G.},
  \bibinfo{author}{Burke, R.}, \bibinfo{author}{Guy, I.},
  \bibinfo{author}{Jannach, D.}, \bibinfo{author}{Kamishima, T.},
  \bibinfo{author}{Krasnodebski, J.}, \& \bibinfo{author}{Pizzato, L.}
  (\bibinfo{year}{2020}).
\newblock \bibinfo{title}{Multistakeholder recommendation: Survey and research
  directions}.
\newblock {\it \bibinfo{journal}{User Modeling and User-Adapted
  Interaction}\/},  {\it \bibinfo{volume}{30}\/}, \bibinfo{pages}{127--158}.
  \DOIprefix\doi{10.1007/s11257-019-09256-1}.
\bibitem[{Adomavicius et~al.(2018)Adomavicius, Bockstedt, Curley \&
  Zhang}]{DBLP:journals/isr/AdomaviciusBCZ18}
\bibinfo{author}{Adomavicius, G.}, \bibinfo{author}{Bockstedt, J.~C.},
  \bibinfo{author}{Curley, S.~P.}, \& \bibinfo{author}{Zhang, J.}
  (\bibinfo{year}{2018}).
\newblock \bibinfo{title}{Effects of online recommendations on consumers'
  willingness to pay}.
\newblock {\it \bibinfo{journal}{Information Systems Research}\/},  {\it
  \bibinfo{volume}{29}\/}, \bibinfo{pages}{84--102}.
\bibitem[{Bauer \& Schedl(2017)}]{Bauer.2017}
\bibinfo{author}{Bauer, C.}, \& \bibinfo{author}{Schedl, M.}
  (\bibinfo{year}{2017}).
\newblock \bibinfo{title}{Introducing surprise and opposition by design in
  recommender systems}.
\newblock In {\it \bibinfo{booktitle}{Proceedings of the 25th Conference on
  User Modeling, Adaptation and Personalization}\/} (pp.
  \bibinfo{pages}{350--353}).
\newblock \DOIprefix\doi{10.1145/3099023.3099099}.
\bibitem[{Bothos et~al.(2015)Bothos, Apostolou \& Mentzas}]{Bothos.2015}
\bibinfo{author}{Bothos, E.}, \bibinfo{author}{Apostolou, D.}, \&
  \bibinfo{author}{Mentzas, G.} (\bibinfo{year}{2015}).
\newblock \bibinfo{title}{Recommender systems for nudging commuters towards
  eco-friendly decisions}.
\newblock {\it \bibinfo{journal}{Intelligent Decision Technologies}\/},  {\it
  \bibinfo{volume}{9}\/}, \bibinfo{pages}{295--306}.
  \DOIprefix\doi{10.3233/IDT-140223}.
\bibitem[{Brown(2019)}]{Brown.2019}
\bibinfo{author}{Brown, C.} (\bibinfo{year}{2019}).
\newblock \bibinfo{title}{Digital nudges for encouraging developer actions}.
\newblock In {\it \bibinfo{booktitle}{IEEE/ACM 41st International Conference on
  Software Engineering: Companion Proceedings (ICSE-Companion)}\/} (pp.
  \bibinfo{pages}{202--205}).
\newblock \DOIprefix\doi{10.1109/ICSE-Companion.2019.00082}.
\bibitem[{Caraban et~al.(2019)Caraban, Karapanos, Gon\c{c}alves \&
  Campos}]{23waystonudge}
\bibinfo{author}{Caraban, A.}, \bibinfo{author}{Karapanos, E.},
  \bibinfo{author}{Gon\c{c}alves, D.}, \& \bibinfo{author}{Campos, P.}
  (\bibinfo{year}{2019}).
\newblock \bibinfo{title}{23 ways to nudge: A review of technology-mediated
  nudging in human-computer interaction}.
\newblock In {\it \bibinfo{booktitle}{Proceedings of the 2019 CHI Conference on
  Human Factors in Computing Systems}\/} (CHI '19) (pp.
  \bibinfo{pages}{1--–15}).
\bibitem[{Cheng \& Langbort(2016)}]{Cheng.2016}
\bibinfo{author}{Cheng, Y.}, \& \bibinfo{author}{Langbort, C.}
  (\bibinfo{year}{2016}).
\newblock \bibinfo{title}{A model of informational nudging in transportation
  networks}.
\newblock In {\it \bibinfo{booktitle}{2016 IEEE 55th Conference on Decision and
  Control (CDC)}\/} (pp. \bibinfo{pages}{7598--7604}).
\newblock \DOIprefix\doi{10.1109/CDC.2016.7799443}.
\bibitem[{Djurica \& Figl(2017)}]{Djurica2017TheEO}
\bibinfo{author}{Djurica, D.}, \& \bibinfo{author}{Figl, K.}
  (\bibinfo{year}{2017}).
\newblock \bibinfo{title}{The effect of digital nudging techniques on
  customers’ product choice and attitudes towards e-commerce sites}.
\newblock In {\it \bibinfo{booktitle}{Proceedings of the 23rd Americas
  Conference on Information Systems (AMCIS)}\/} (pp. \bibinfo{pages}{1--5}).
\bibitem[{Dolan et~al.(2012)Dolan, Hallsworth, Halpern, King, Metcalfe \&
  Vlaev}]{Dolan.2012}
\bibinfo{author}{Dolan, P.}, \bibinfo{author}{Hallsworth, M.},
  \bibinfo{author}{Halpern, D.}, \bibinfo{author}{King, D.},
  \bibinfo{author}{Metcalfe, R.}, \& \bibinfo{author}{Vlaev, I.}
  (\bibinfo{year}{2012}).
\newblock \bibinfo{title}{Influencing behaviour: {T}he mindspace way}.
\newblock {\it \bibinfo{journal}{Journal of Economic Psychology}\/},  {\it
  \bibinfo{volume}{33}\/}, \bibinfo{pages}{264--277}.
  \DOIprefix\doi{10.1016/j.joep.2011.10.009}.
\bibitem[{Dolan et~al.(2010)Dolan, Hallsworth, Halpern, King \&
  Vlaev}]{Mindspace2010}
\bibinfo{author}{Dolan, P.}, \bibinfo{author}{Hallsworth, M.},
  \bibinfo{author}{Halpern, D.}, \bibinfo{author}{King, D.}, \&
  \bibinfo{author}{Vlaev, I.} (\bibinfo{year}{2010}).
\newblock \bibinfo{title}{Mindspace: {I}nfluencing behaviour for public
  policy}.
\bibitem[{{E. Bothos} et~al.(2016){E. Bothos}, {D. Apostolou} \& {G.
  Mentzas}}]{E.Bothos.2016}
\bibinfo{author}{{E. Bothos}}, \bibinfo{author}{{D. Apostolou}}, \&
  \bibinfo{author}{{G. Mentzas}} (\bibinfo{year}{2016}).
\newblock \bibinfo{title}{A recommender for persuasive messages in route
  planning applications}.
\newblock {\it \bibinfo{journal}{2016 7th International Conference on
  Information, Intelligence, Systems and Applications (IISA)}\/},  (pp.
  \bibinfo{pages}{1--5}). \DOIprefix\doi{10.1109/IISA.2016.7785399}.
\bibitem[{Elsweiler et~al.(2017)Elsweiler, Trattner \& Harvey}]{Elsweiler.2017}
\bibinfo{author}{Elsweiler, D.}, \bibinfo{author}{Trattner, C.}, \&
  \bibinfo{author}{Harvey, M.} (\bibinfo{year}{2017}).
\newblock \bibinfo{title}{Exploiting food choice biases for healthier recipe
  recommendation}.
\newblock In {\it \bibinfo{booktitle}{Proceedings of the 40th International ACM
  SIGIR Conference on Research and Development in Information Retrieval}\/}
  (pp. \bibinfo{pages}{575--584}).
\newblock \DOIprefix\doi{10.1145/3077136.3080826}.
\bibitem[{Esposito et~al.(2017)Esposito, Hern{\'a}ndez, {van Bavel} \&
  Vila}]{Esposito.2017}
\bibinfo{author}{Esposito, G.}, \bibinfo{author}{Hern{\'a}ndez, P.},
  \bibinfo{author}{{van Bavel}, R.}, \& \bibinfo{author}{Vila, J.}
  (\bibinfo{year}{2017}).
\newblock \bibinfo{title}{Nudging to prevent the purchase of incompatible
  digital products online: An experimental study}.
\newblock {\it \bibinfo{journal}{PLOS ONE}\/},  {\it \bibinfo{volume}{12}\/},
  \bibinfo{pages}{1--15}. \DOIprefix\doi{10.1371/journal.pone.0173333}.
\bibitem[{Felfernig et~al.(2007)Felfernig, Friedrich, Gula, Hitz, Kruggel,
  Leitner, Melcher, Riepan, Strauss, Teppan et~al.}]{Felfernig.2007}
\bibinfo{author}{Felfernig, A.}, \bibinfo{author}{Friedrich, G.},
  \bibinfo{author}{Gula, B.}, \bibinfo{author}{Hitz, M.},
  \bibinfo{author}{Kruggel, T.}, \bibinfo{author}{Leitner, G.},
  \bibinfo{author}{Melcher, R.}, \bibinfo{author}{Riepan, D.},
  \bibinfo{author}{Strauss, S.}, \bibinfo{author}{Teppan, E.} et~al.
  (\bibinfo{year}{2007}).
\newblock \bibinfo{title}{Persuasive recommendation: Serial position effects in
  knowledge-based recommender systems}.
\newblock In {\it \bibinfo{booktitle}{Persuasive Technology}\/} (pp.
  \bibinfo{pages}{283--294}).
\bibitem[{Felfernig et~al.(2008)Felfernig, Gula, Leitner, Maier, Melcher \&
  Teppan}]{Felfernig.2008}
\bibinfo{author}{Felfernig, A.}, \bibinfo{author}{Gula, B.},
  \bibinfo{author}{Leitner, G.}, \bibinfo{author}{Maier, M.},
  \bibinfo{author}{Melcher, R.}, \& \bibinfo{author}{Teppan, E.}
  (\bibinfo{year}{2008}).
\newblock \bibinfo{title}{Persuasion in knowledge-based recommendation}.
\newblock In {\it \bibinfo{booktitle}{Persuasive technology}\/} Lecture Notes
  in Computer Science (pp. \bibinfo{pages}{71--82}).
\newblock \DOIprefix\doi{10.1007/978-3-540-68504-3/7}.
\bibitem[{Fogg(2002)}]{FoggPresuasion}
\bibinfo{author}{Fogg, B.} (\bibinfo{year}{2002}).
\newblock \bibinfo{title}{Persuasive technology: {U}sing computers to change
  what we think and do}.
\newblock {\it \bibinfo{journal}{Ubiquity}\/},  {\it \bibinfo{volume}{2002}\/},
  \bibinfo{pages}{2}. \DOIprefix\doi{10.1145/763955.763957}.
\bibitem[{Forwood et~al.(2015)Forwood, Ahern, Marteau \& Jebb}]{Forwood.2015}
\bibinfo{author}{Forwood, S.~E.}, \bibinfo{author}{Ahern, A.~L.},
  \bibinfo{author}{Marteau, T.~M.}, \& \bibinfo{author}{Jebb, S.~A.}
  (\bibinfo{year}{2015}).
\newblock \bibinfo{title}{Offering within-category food swaps to reduce energy
  density of food purchases: {A} study using an experimental online
  supermarket}.
\newblock {\it \bibinfo{journal}{International Journal of Behavioral Nutrition
  and Physical Activity}\/},  {\it \bibinfo{volume}{12}\/},
  \bibinfo{pages}{85}. \DOIprefix\doi{10.1186/s12966-015-0241-1}.
\bibitem[{Gena et~al.(2019)Gena, Grillo, Lieto, Mattutino \&
  Vernero}]{Gena.2019}
\bibinfo{author}{Gena, C.}, \bibinfo{author}{Grillo, P.},
  \bibinfo{author}{Lieto, A.}, \bibinfo{author}{Mattutino, C.}, \&
  \bibinfo{author}{Vernero, F.} (\bibinfo{year}{2019}).
\newblock \bibinfo{title}{When personalization is not an option: An in-the-wild
  study on persuasive news recommendation}.
\newblock {\it \bibinfo{journal}{Information}\/},  {\it
  \bibinfo{volume}{10}\/}, \bibinfo{pages}{300}.
\bibitem[{Gomez-Uribe \& Hunt(2015)}]{Gomez-Uribe:2015:NRS:2869770.2843948}
\bibinfo{author}{Gomez-Uribe, C.~A.}, \& \bibinfo{author}{Hunt, N.}
  (\bibinfo{year}{2015}).
\newblock \bibinfo{title}{The {Netflix} recommender system: Algorithms,
  business value, and innovation}.
\newblock {\it \bibinfo{journal}{{ACM} Transactions on Management Information
  Systems}\/},  {\it \bibinfo{volume}{6}\/}, \bibinfo{pages}{1--19}.
\bibitem[{Guers et~al.(2013)Guers, Langbort \& Work}]{Guers.2013}
\bibinfo{author}{Guers, R.}, \bibinfo{author}{Langbort, C.}, \&
  \bibinfo{author}{Work, D.} (\bibinfo{year}{2013}).
\newblock \bibinfo{title}{On informational nudging and control of payoff-based
  learning}.
\newblock {\it \bibinfo{journal}{IFAC Proceedings Volumes}\/},  {\it
  \bibinfo{volume}{46}\/}, \bibinfo{pages}{69--74}.
  \DOIprefix\doi{10.3182/20130925-2-DE-4044.00037}.
\bibitem[{He \& McAuley(2016)}]{he2016vbpr}
\bibinfo{author}{He, R.}, \& \bibinfo{author}{McAuley, J.}
  (\bibinfo{year}{2016}).
\newblock \bibinfo{title}{Vbpr: {V}isual bayesian personalized ranking from
  implicit feedback}.
\newblock In {\it \bibinfo{booktitle}{Thirtieth AAAI Conference on Artificial
  Intelligence}\/} (p. \bibinfo{pages}{144–150}).
\bibitem[{Huhn et~al.(2018)Huhn, Ferreira, Freitas \& Leão}]{Huhn.2018}
\bibinfo{author}{Huhn, R.}, \bibinfo{author}{Ferreira, J.},
  \bibinfo{author}{Freitas, A.}, \& \bibinfo{author}{Leão, F.}
  (\bibinfo{year}{2018}).
\newblock \bibinfo{title}{The effects of social media opinion leaders’
  recommendations on followers’ intention to buy}.
\newblock {\it \bibinfo{journal}{Review of Business Management}\/},  {\it
  \bibinfo{volume}{20}\/}, \bibinfo{pages}{57--73}.
  \DOIprefix\doi{10.7819/rbgn.v20i1.3678}.
\bibitem[{Hummel \& Maedche(2019)}]{HUMMEL201947}
\bibinfo{author}{Hummel, D.}, \& \bibinfo{author}{Maedche, A.}
  (\bibinfo{year}{2019}).
\newblock \bibinfo{title}{How effective is nudging? {A} quantitative review on
  the effect sizes and limits of empirical nudging studies}.
\newblock {\it \bibinfo{journal}{Journal of Behavioral and Experimental
  Economics}\/},  {\it \bibinfo{volume}{80}\/}, \bibinfo{pages}{47 -- 58}.
\bibitem[{Jannach(2004)}]{Jannach2004}
\bibinfo{author}{Jannach, D.} (\bibinfo{year}{2004}).
\newblock \bibinfo{title}{{ADVISOR} {SUITE} - {A} knowledge-based sales
  advisory-system}.
\newblock In {\it \bibinfo{booktitle}{Proceedings of the 16th Eureopean
  Conference on Artificial Intelligence ({ECAI 2004})}\/} (pp.
  \bibinfo{pages}{720--724}).
\bibitem[{Jannach \& Adomavicius(2016)}]{JannachAdomavicius2016purpose}
\bibinfo{author}{Jannach, D.}, \& \bibinfo{author}{Adomavicius, G.}
  (\bibinfo{year}{2016}).
\newblock \bibinfo{title}{Recommendations with a purpose}.
\newblock In {\it \bibinfo{booktitle}{Proceedings of the 10th ACM Conference on
  Recommender Systems}\/} (RecSys '16) (pp. \bibinfo{pages}{7--10}).
\newblock \DOIprefix\doi{10.1145/2959100.2959186}.
\bibitem[{Jannach \& Adomavicius(2017)}]{JannachAdomaviciusVAMS2017}
\bibinfo{author}{Jannach, D.}, \& \bibinfo{author}{Adomavicius, G.}
  (\bibinfo{year}{2017}).
\newblock \bibinfo{title}{Price and profit awareness in recommender systems}.
\newblock In {\it \bibinfo{booktitle}{Proceedings of the ACM RecSys 2017
  Workshop on Value-Aware and Multi-Stakeholder Recommendation}\/}
  (p.~\bibinfo{pages}{5}).
\bibitem[{Jannach \& Jugovac(2019)}]{jannachjugovactmis2019}
\bibinfo{author}{Jannach, D.}, \& \bibinfo{author}{Jugovac, M.}
  (\bibinfo{year}{2019}).
\newblock \bibinfo{title}{Measuring the business value of recommender systems}.
\newblock {\it \bibinfo{journal}{{ACM} Transactions on Management Information
  Systems}\/},  {\it \bibinfo{volume}{10}\/}, \bibinfo{pages}{23}.
\bibitem[{Johnson et~al.(2012{\natexlab{a}})Johnson, Shu, Dellaert, Fox,
  Goldstein, H{\"a}ubl, Larrick, Payne, Peters, Schkade, Wansink \&
  Weber}]{Johnson.2012}
\bibinfo{author}{Johnson, E.}, \bibinfo{author}{Shu, S.},
  \bibinfo{author}{Dellaert, B.}, \bibinfo{author}{Fox, C.},
  \bibinfo{author}{Goldstein, D.}, \bibinfo{author}{H{\"a}ubl, G.},
  \bibinfo{author}{Larrick, R.}, \bibinfo{author}{Payne, J.},
  \bibinfo{author}{Peters, E.}, \bibinfo{author}{Schkade, D.},
  \bibinfo{author}{Wansink, B.}, \& \bibinfo{author}{Weber, E.}
  (\bibinfo{year}{2012}{\natexlab{a}}).
\newblock \bibinfo{title}{Beyond nudges: Tools of a choice architecture}.
\newblock {\it \bibinfo{journal}{Marketing Letters}\/},  {\it
  \bibinfo{volume}{23}\/}, \bibinfo{pages}{487--504}.
  \DOIprefix\doi{10.1007/s11002-012-9186-1}.
\bibitem[{Johnson et~al.(2012{\natexlab{b}})Johnson, Shu, Dellaert, Fox,
  Goldstein, H\"{a}ubl, Larrick, Payne, Peters, Schkade, Wansink \&
  Weber}]{BeyondNudges202}
\bibinfo{author}{Johnson, E.~J.}, \bibinfo{author}{Shu, S.~B.},
  \bibinfo{author}{Dellaert, B. G.~C.}, \bibinfo{author}{Fox, C.},
  \bibinfo{author}{Goldstein, D.~G.}, \bibinfo{author}{H\"{a}ubl, G.},
  \bibinfo{author}{Larrick, R.~P.}, \bibinfo{author}{Payne, J.~W.},
  \bibinfo{author}{Peters, E.}, \bibinfo{author}{Schkade, D.},
  \bibinfo{author}{Wansink, B.}, \& \bibinfo{author}{Weber, E.~U.}
  (\bibinfo{year}{2012}{\natexlab{b}}).
\newblock \bibinfo{title}{Beyond nudges: Tools of a choice architecture}.
\newblock {\it \bibinfo{journal}{Marketing Letters: A Journal of Research in
  Marketing}\/},  {\it \bibinfo{volume}{23}\/}, \bibinfo{pages}{487--504}.
\bibitem[{Jung et~al.(2018)Jung, Erdfelder \& Glaser}]{Jung.2018}
\bibinfo{author}{Jung, D.}, \bibinfo{author}{Erdfelder, E.}, \&
  \bibinfo{author}{Glaser, F.} (\bibinfo{year}{2018}).
\newblock \bibinfo{title}{Nudged to win: Designing robo-advisory to overcome
  decision inertia}.
\newblock In {\it \bibinfo{booktitle}{Proceedings of the 26th European
  Conference on Information Systems (ECIS 2018)}\/} (p.~\bibinfo{pages}{12}).
\bibitem[{Kahneman(2011)}]{kahneman2011thinking}
\bibinfo{author}{Kahneman, D.} (\bibinfo{year}{2011}).
\newblock {\it \bibinfo{title}{Thinking, fast and slow}\/}.
\newblock \bibinfo{publisher}{Macmillan}.
\bibitem[{Kamehkhosh et~al.(2019)Kamehkhosh, Bonnin \&
  Jannach}]{KamehkhoschBonninJannach19}
\bibinfo{author}{Kamehkhosh, I.}, \bibinfo{author}{Bonnin, G.}, \&
  \bibinfo{author}{Jannach, D.} (\bibinfo{year}{2019}).
\newblock \bibinfo{title}{Effects of recommendations on the playlist creation
  behavior of users}.
\newblock {\it \bibinfo{journal}{User Modeling and User-Adapted
  Interaction}\/},  {\it \bibinfo{volume}{30}\/}, \bibinfo{pages}{285–322}.
\bibitem[{Kammerer \& Gerjets(2014)}]{Kammerer.2014}
\bibinfo{author}{Kammerer, Y.}, \& \bibinfo{author}{Gerjets, P.}
  (\bibinfo{year}{2014}).
\newblock \bibinfo{title}{The role of search result position and source
  trustworthiness in the selection of web search results when using a list or a
  grid interface}.
\newblock {\it \bibinfo{journal}{International Journal of Human-Computer
  Interaction}\/},  {\it \bibinfo{volume}{30}\/}, \bibinfo{pages}{177--191}.
  \DOIprefix\doi{10.1080/10447318.2013.846790}.
\bibitem[{Karlsen \& Andersen(2019)}]{Recommendationswithanudge2019}
\bibinfo{author}{Karlsen, R.}, \& \bibinfo{author}{Andersen, A.}
  (\bibinfo{year}{2019}).
\newblock \bibinfo{title}{Recommendations with a nudge}.
\newblock {\it \bibinfo{journal}{Technologies}\/},  {\it
  \bibinfo{volume}{7}\/}, \bibinfo{pages}{45}.
  \DOIprefix\doi{10.3390/technologies7020045}.
\bibitem[{Karvonen et~al.(2010)Karvonen, Shibasaki, Nunes, Kaur \&
  Immonen}]{Karvonen.2010}
\bibinfo{author}{Karvonen, K.}, \bibinfo{author}{Shibasaki, S.},
  \bibinfo{author}{Nunes, S.}, \bibinfo{author}{Kaur, P.}, \&
  \bibinfo{author}{Immonen, O.} (\bibinfo{year}{2010}).
\newblock \bibinfo{title}{Visual nudges for enhancing the use and produce of
  reputation information}.
\newblock In {\it \bibinfo{booktitle}{Proceedings of the Workshop on
  User-Centric Evaluation of Recommender Systems and Their Interfaces at the
  4th Conference on Recommender Systems (RecSys’10)}\/} (pp.
  \bibinfo{pages}{1--8}).
\newblock volume \bibinfo{volume}{612}.
\bibitem[{Köcher et~al.(2019)Köcher, Jugovac, Jannach \&
  Holzmüller}]{KoecherEtAl2018}
\bibinfo{author}{Köcher, S.}, \bibinfo{author}{Jugovac, M.},
  \bibinfo{author}{Jannach, D.}, \& \bibinfo{author}{Holzmüller, H.}
  (\bibinfo{year}{2019}).
\newblock \bibinfo{title}{New hidden persuaders: An investigation of
  attribute-level anchoring effects of product recommendations}.
\newblock {\it \bibinfo{journal}{Journal of Retailing}\/},  {\it
  \bibinfo{volume}{95}\/}, \bibinfo{pages}{24--41}.
\bibitem[{Kissmer et~al.(2018)Kissmer, Kroll \& Stieglitz}]{Kissmer.2018}
\bibinfo{author}{Kissmer, T.}, \bibinfo{author}{Kroll, T.}, \&
  \bibinfo{author}{Stieglitz, S.} (\bibinfo{year}{2018}).
\newblock \bibinfo{title}{Enterprise digital nudging: Between adoption gain and
  unintended rejection}.
\newblock In {\it \bibinfo{booktitle}{Proceedings of the 24th Americas
  Conference on Information (AMCIS)}\/} (p.~\bibinfo{pages}{5}).
\bibitem[{Knijnenburg(2014)}]{Knijnenburg.2014}
\bibinfo{author}{Knijnenburg, B.~P.} (\bibinfo{year}{2014}).
\newblock \bibinfo{title}{Information disclosure profiles for segmentation and
  recommendation}.
\newblock In {\it \bibinfo{booktitle}{SOUPS2014 Workshop on Privacy Personas
  and Segmentation}\/} (pp. \bibinfo{pages}{1--4}).
\bibitem[{Lee \& Hosanagar(2019)}]{LeeHosanagar2018}
\bibinfo{author}{Lee, D.}, \& \bibinfo{author}{Hosanagar, K.}
  (\bibinfo{year}{2019}).
\newblock \bibinfo{title}{How do recommender systems affect sales diversity?
  {A} cross-category investigation via randomized field experiment}.
\newblock {\it \bibinfo{journal}{Information Systems Research}\/},  {\it
  \bibinfo{volume}{30}\/}, \bibinfo{pages}{239--259}.
\bibitem[{Lee et~al.(2011)Lee, Kiesler \& Forlizzi}]{Lee.2011}
\bibinfo{author}{Lee, M.~K.}, \bibinfo{author}{Kiesler, S.}, \&
  \bibinfo{author}{Forlizzi, J.} (\bibinfo{year}{2011}).
\newblock \bibinfo{title}{Mining behavioral economics to design persuasive
  technology for healthy choices}.
\newblock {\it \bibinfo{journal}{Conference on Human Factors in Computing
  Systems - Proceedings}\/},  (pp. \bibinfo{pages}{325--334}).
  \DOIprefix\doi{10.1145/1978942.1978989}.
\bibitem[{Leitner(2015)}]{Leitner.2015}
\bibinfo{author}{Leitner, G.} (\bibinfo{year}{2015}).
\newblock \bibinfo{title}{Psyrec: {P}sychological concepts to enhance the
  interaction with recommender systems}.
\newblock In {\it \bibinfo{booktitle}{CEUR Workshop Proceedings}\/} (pp.
  \bibinfo{pages}{37--44}).
\newblock volume \bibinfo{volume}{1349}.
\bibitem[{Leonard et~al.(2008)Leonard, Thaler \& Sunstein}]{thaler2008nudge}
\bibinfo{author}{Leonard, T.~C.}, \bibinfo{author}{Thaler, R.~H.}, \&
  \bibinfo{author}{Sunstein, C.~R.} (\bibinfo{year}{2008}).
\newblock \bibinfo{title}{Nudge: Improving decisions about health, wealth, and
  happiness}.
\newblock {\it \bibinfo{journal}{Constitutional Political Economy}\/},  {\it
  \bibinfo{volume}{19}\/}, \bibinfo{pages}{356--360}.
  \DOIprefix\doi{10.1007/s10602-008-9056-2}.
\bibitem[{Loewenstein et~al.(2015)Loewenstein, Bryce, Hagmann \&
  Rajpal}]{Loewenstein.2015}
\bibinfo{author}{Loewenstein, G.}, \bibinfo{author}{Bryce, C.},
  \bibinfo{author}{Hagmann, D.}, \& \bibinfo{author}{Rajpal, S.}
  (\bibinfo{year}{2015}).
\newblock \bibinfo{title}{Warning: You are about to be nudged}.
\newblock {\it \bibinfo{journal}{Behavioral Science {\&} Policy}\/},  {\it
  \bibinfo{volume}{1}\/}, \bibinfo{pages}{35--42}.
\bibitem[{Loewenstein \& Chater(2017)}]{Loewenstein.2017}
\bibinfo{author}{Loewenstein, G.}, \& \bibinfo{author}{Chater, N.}
  (\bibinfo{year}{2017}).
\newblock \bibinfo{title}{Putting nudges in perspective}.
\newblock {\it \bibinfo{journal}{Behavioural Public Policy}\/},  {\it
  \bibinfo{volume}{1}\/}, \bibinfo{pages}{26--53}.
  \DOIprefix\doi{10.1017/bpp.2016.7}.
\bibitem[{Meske \& Potthoff(2017)}]{DBLP:conf/ecis/MeskeP17}
\bibinfo{author}{Meske, C.}, \& \bibinfo{author}{Potthoff, T.}
  (\bibinfo{year}{2017}).
\newblock \bibinfo{title}{The {DINU}-{M}odel - {A} process model for the design
  of nudges}.
\newblock {\it \bibinfo{journal}{25th European Conference on Information
  Systems, ({ECIS} 2017)}\/},  (pp. \bibinfo{pages}{2587--2597}).
\bibitem[{Miller(2019)}]{Miller.2019}
\bibinfo{author}{Miller, T.} (\bibinfo{year}{2019}).
\newblock \bibinfo{title}{Explanation in artificial intelligence: Insights from
  the social sciences}.
\newblock {\it \bibinfo{journal}{Artificial Intelligence}\/},  {\it
  \bibinfo{volume}{267}\/}, \bibinfo{pages}{1--38}.
  \DOIprefix\doi{10.1016/j.artint.2018.07.007}.
\bibitem[{{Mimmi Castmo} \& {Rebecka Persson}(2018)}]{MimmiCastmo.2018}
\bibinfo{author}{{Mimmi Castmo}}, \& \bibinfo{author}{{Rebecka Persson}}
  (\bibinfo{year}{2018}).
\newblock \bibinfo{title}{The alliance of digital nudging {\&} persuasive
  design: The complementary nature of the design strategies}.
\newblock {\it \bibinfo{journal}{undefined}\/},  (p.~\bibinfo{pages}{49}).
\bibitem[{Mirsch et~al.(2018{\natexlab{a}})Mirsch, Jung, Rieder \&
  Lehrer}]{Mirsch.2018}
\bibinfo{author}{Mirsch, T.}, \bibinfo{author}{Jung, R.},
  \bibinfo{author}{Rieder, A.}, \& \bibinfo{author}{Lehrer, C.}
  (\bibinfo{year}{2018}{\natexlab{a}}).
\newblock \bibinfo{title}{Mit {D}igital {N}udging {N}utzererlebnisse verbessern
  und den {U}nternehmenserfolg steigern}.
\newblock {\it \bibinfo{journal}{Controlling}\/},  {\it
  \bibinfo{volume}{30}\/}, \bibinfo{pages}{12--18}.
  \DOIprefix\doi{10.15358/0935-0381-2018-5-12}.
\bibitem[{Mirsch et~al.(2017)Mirsch, Lehrer \& Jung}]{DBLP:conf/wi/MirschLJ17}
\bibinfo{author}{Mirsch, T.}, \bibinfo{author}{Lehrer, C.}, \&
  \bibinfo{author}{Jung, R.} (\bibinfo{year}{2017}).
\newblock \bibinfo{title}{Digital nudging: Altering user behavior in digital
  environments}.
\newblock In {\it \bibinfo{booktitle}{Towards Thought Leadership in Digital
  Transformation: 13. Internationale Tagung Wirtschaftsinformatik}\/} (pp.
  \bibinfo{pages}{634--648}).
\bibitem[{Mirsch et~al.(2018{\natexlab{b}})Mirsch, Lehrer \&
  Jung}]{Mirsch.2018b}
\bibinfo{author}{Mirsch, T.}, \bibinfo{author}{Lehrer, C.}, \&
  \bibinfo{author}{Jung, R.} (\bibinfo{year}{2018}{\natexlab{b}}).
\newblock \bibinfo{title}{Making digital nudging applicable: {T}he digital
  nudge design method}.
\newblock In {\it \bibinfo{booktitle}{Proceedings of the 39th International
  Conference on Information Systems (ICIS)}\/} (p.~\bibinfo{pages}{19}).
\bibitem[{M\"{u}nscher et~al.(2016)M\"{u}nscher, Vetter \&
  Scheuerle}]{Muenscher2016}
\bibinfo{author}{M\"{u}nscher, R.}, \bibinfo{author}{Vetter, M.}, \&
  \bibinfo{author}{Scheuerle, T.} (\bibinfo{year}{2016}).
\newblock \bibinfo{title}{A review and taxonomy of choice architecture
  techniques}.
\newblock {\it \bibinfo{journal}{Journal of Behavioral Decision Making}\/},
  {\it \bibinfo{volume}{29}\/}, \bibinfo{pages}{511--524}.
\bibitem[{Rook et~al.(2020)Rook, Sabic \& Zanker}]{DBLP:journals/jiis/RookSZ20}
\bibinfo{author}{Rook, L.}, \bibinfo{author}{Sabic, A.}, \&
  \bibinfo{author}{Zanker, M.} (\bibinfo{year}{2020}).
\newblock \bibinfo{title}{Engagement in proactive recommendations}.
\newblock {\it \bibinfo{journal}{Journal of Intelligent Information
  Systems}\/},  {\it \bibinfo{volume}{54}\/}, \bibinfo{pages}{79--100}.
\bibitem[{Sch{\"a}r \& Stanoevska-Slabeva(2019)}]{Schaer2019}
\bibinfo{author}{Sch{\"a}r, A.}, \& \bibinfo{author}{Stanoevska-Slabeva, K.}
  (\bibinfo{year}{2019}).
\newblock \bibinfo{title}{Application of digital nudging in customer journeys?
  {A} systematic literature review}.
\newblock In {\it \bibinfo{booktitle}{Proceedings of the 25th Americas
  Conference on Information Systems (AMCIS)}\/} (pp. \bibinfo{pages}{1--10}).
\bibitem[{Schneider et~al.(2018)Schneider, Weinmann \& vom
  Brocke}]{DBLP:journals/cacm/SchneiderWB18}
\bibinfo{author}{Schneider, C.}, \bibinfo{author}{Weinmann, M.}, \&
  \bibinfo{author}{vom Brocke, J.} (\bibinfo{year}{2018}).
\newblock \bibinfo{title}{Digital nudging: {G}uiding online user choices
  through interface design}.
\newblock {\it \bibinfo{journal}{Communications of the {ACM}}\/},  {\it
  \bibinfo{volume}{61}\/}, \bibinfo{pages}{67--73}.
\bibitem[{Sunstein(2014)}]{NudgingAVeryShortGuid42014}
\bibinfo{author}{Sunstein, C.~R.} (\bibinfo{year}{2014}).
\newblock \bibinfo{title}{Nudging: A very short guide}.
\newblock {\it \bibinfo{journal}{Journal of Consumer Policy}\/},  {\it
  \bibinfo{volume}{37}\/}, \bibinfo{pages}{583--588}.
\bibitem[{Sunstein(2015)}]{Sunstein2015NudgingAC}
\bibinfo{author}{Sunstein, C.~R.} (\bibinfo{year}{2015}).
\newblock \bibinfo{title}{Nudging and choice architecture: Ethical
  considerations}.
\newblock {\it \bibinfo{journal}{Yale Journal on Regulation, (2015)}\/}, .
\bibitem[{Sunstein(2016)}]{SunsteinCouncil2016}
\bibinfo{author}{Sunstein, C.~R.} (\bibinfo{year}{2016}).
\newblock \bibinfo{title}{The council of psychological advisers}.
\newblock {\it \bibinfo{journal}{Annual Review of Psychology}\/},  {\it
  \bibinfo{volume}{67}\/}, \bibinfo{pages}{713--–737}.
\bibitem[{Szaszi et~al.(2018)Szaszi, Palinkas, Palfi, Szollosi \&
  Aczel}]{Szaszi.2018}
\bibinfo{author}{Szaszi, B.}, \bibinfo{author}{Palinkas, A.},
  \bibinfo{author}{Palfi, B.}, \bibinfo{author}{Szollosi, A.}, \&
  \bibinfo{author}{Aczel, B.} (\bibinfo{year}{2018}).
\newblock \bibinfo{title}{A systematic scoping review of the choice
  architecture movement: Toward understanding when and why nudges work}.
\newblock {\it \bibinfo{journal}{Journal of Behavioral Decision Making}\/},
  {\it \bibinfo{volume}{31}\/}, \bibinfo{pages}{355--366}.
  \DOIprefix\doi{10.1002/bdm.2035}.
\bibitem[{Teppan \& Felfernig(2009)}]{Teppan.2009}
\bibinfo{author}{Teppan, E.~C.}, \& \bibinfo{author}{Felfernig, A.}
  (\bibinfo{year}{2009}).
\newblock \bibinfo{title}{Asymmetric dominance- and compromise effects in the
  financial services domain}.
\newblock In {\it \bibinfo{booktitle}{2009 IEEE Conference on Commerce and
  Enterprise Computing}\/} (pp. \bibinfo{pages}{57--64}).
\newblock \DOIprefix\doi{10.1109/CEC.2009.69}.
\bibitem[{Thaler(2018)}]{Thaler.2018}
\bibinfo{author}{Thaler, R.~H.} (\bibinfo{year}{2018}).
\newblock \bibinfo{title}{From cashews to nudges: The evolution of behavioral
  economics}.
\newblock {\it \bibinfo{journal}{American Economic Review}\/},  {\it
  \bibinfo{volume}{108}\/}, \bibinfo{pages}{1265--1287}.
  \DOIprefix\doi{10.1257/aer.108.6.1265}.
\bibitem[{Thaler et~al.(2012)Thaler, Sunstein \& Balz}]{ChoiceArchitecture2014}
\bibinfo{author}{Thaler, R.~H.}, \bibinfo{author}{Sunstein, C.~R.}, \&
  \bibinfo{author}{Balz, J.~P.} (\bibinfo{year}{2012}).
\newblock \bibinfo{title}{Choice architecture}.
\newblock {\it \bibinfo{journal}{The Behavioral Foundations of Public
  Policy}\/},  (pp. \bibinfo{pages}{428--439}).
\bibitem[{Theocharous et~al.(2019)Theocharous, Healey, Mahadevan \&
  Saad}]{Theocharous.2019}
\bibinfo{author}{Theocharous, G.}, \bibinfo{author}{Healey, J.},
  \bibinfo{author}{Mahadevan, S.}, \& \bibinfo{author}{Saad, M.}
  (\bibinfo{year}{2019}).
\newblock \bibinfo{title}{Personalizing with human cognitive biases}.
\newblock In {\it \bibinfo{booktitle}{Proceedings of the 27th ACM Conference on
  User Modeling, Adaptation and Personalization}\/} (pp.
  \bibinfo{pages}{13--17}).
\newblock \DOIprefix\doi{10.1145/3314183.3323453}.
\bibitem[{Turland et~al.(2015)Turland, Coventry, Jeske, Briggs \& {van
  Moorsel}}]{Turland.2015}
\bibinfo{author}{Turland, J.}, \bibinfo{author}{Coventry, L.},
  \bibinfo{author}{Jeske, D.}, \bibinfo{author}{Briggs, P.}, \&
  \bibinfo{author}{{van Moorsel}, A.} (\bibinfo{year}{2015}).
\newblock \bibinfo{title}{Nudging towards security: Developing an application
  for wireless network selection for android phones}.
\newblock {\it \bibinfo{journal}{Proceedings of the 2015 British HCI
  Conference}\/},  (pp. \bibinfo{pages}{193--201}).
  \DOIprefix\doi{10.1145/2783446.2783588}.
\bibitem[{Tversky \& Kahneman(1974{\natexlab{a}})}]{Tversky1124}
\bibinfo{author}{Tversky, A.}, \& \bibinfo{author}{Kahneman, D.}
  (\bibinfo{year}{1974}{\natexlab{a}}).
\newblock \bibinfo{title}{Judgment under uncertainty: Heuristics and biases}.
\newblock {\it \bibinfo{journal}{Science}\/},  {\it \bibinfo{volume}{185}\/},
  \bibinfo{pages}{1124--1131}.
\bibitem[{Tversky \& Kahneman(1974{\natexlab{b}})}]{Tversky.1974}
\bibinfo{author}{Tversky, A.}, \& \bibinfo{author}{Kahneman, D.}
  (\bibinfo{year}{1974}{\natexlab{b}}).
\newblock \bibinfo{title}{Judgment under uncertainty: Heuristics and biases}.
\newblock {\it \bibinfo{journal}{Science}\/},  {\it \bibinfo{volume}{185}\/},
  \bibinfo{pages}{1124--1131}. \DOIprefix\doi{10.1126/science.185.4157.1124}.
\bibitem[{Verma et~al.(2018)Verma, Wadhwa, Singh, Beniwal, Kaushal \&
  Kumaraguru}]{Verma.2018}
\bibinfo{author}{Verma, A.}, \bibinfo{author}{Wadhwa, A.},
  \bibinfo{author}{Singh, N.}, \bibinfo{author}{Beniwal, S.},
  \bibinfo{author}{Kaushal, R.}, \& \bibinfo{author}{Kumaraguru, P.}
  (\bibinfo{year}{2018}).
\newblock \bibinfo{title}{Followee management: {H}elping users follow the right
  users on online social media}.
\newblock In {\it \bibinfo{booktitle}{IEEE/ACM International Conference on
  Advances in Social Networks Analysis and Mining (ASONAM)}\/} (pp.
  \bibinfo{pages}{1286--1290}).
\bibitem[{Wang et~al.(2019)Wang, Shi, Kim, Oh, Yang, Zhang \& Yu}]{Wang.2019}
\bibinfo{author}{Wang, X.}, \bibinfo{author}{Shi, W.}, \bibinfo{author}{Kim,
  R.}, \bibinfo{author}{Oh, Y.}, \bibinfo{author}{Yang, S.},
  \bibinfo{author}{Zhang, J.}, \& \bibinfo{author}{Yu, Z.}
  (\bibinfo{year}{2019}).
\newblock \bibinfo{title}{Persuasion for good: Towards a personalized
  persuasive dialogue system for social good}.
\newblock In {\it \bibinfo{booktitle}{Proceedings of the 57th Annual Meeting of
  the Association for Computational Linguistics}\/} (pp.
  \bibinfo{pages}{5635--5649}).
\newblock \DOIprefix\doi{10.18653/v1/P19-1566}.
\bibitem[{Weinmann et~al.(2016)Weinmann, Schneider \& vom
  Brocke}]{DBLP:journals/bise/WeinmannSB16}
\bibinfo{author}{Weinmann, M.}, \bibinfo{author}{Schneider, C.}, \&
  \bibinfo{author}{vom Brocke, J.} (\bibinfo{year}{2016}).
\newblock \bibinfo{title}{Digital nudging}.
\newblock {\it \bibinfo{journal}{Business \& Information Systems
  Engineering}\/},  {\it \bibinfo{volume}{58}\/}, \bibinfo{pages}{433--436}.
\bibitem[{Yoo et~al.(2013)Yoo, Gretzel \& Zanker}]{YooPersuasiveRS2013}
\bibinfo{author}{Yoo, K.-H.}, \bibinfo{author}{Gretzel, U.}, \&
  \bibinfo{author}{Zanker, M.} (\bibinfo{year}{2013}).
\newblock {\it \bibinfo{title}{Persuasive Recommender Systems -- Conceptual
  Background and Implications}\/}.
\newblock \bibinfo{publisher}{Springer Science \& Business Media}.
\bibitem[{Zhang et~al.(2015)Zhang, Kannan \& Shanthikumar}]{Zhang.2015}
\bibinfo{author}{Zhang, N.}, \bibinfo{author}{Kannan, K.~N.}, \&
  \bibinfo{author}{Shanthikumar, J.~G.} (\bibinfo{year}{2015}).
\newblock \bibinfo{title}{A recommender system to nudge customers in a capacity
  constrained supply chain}.
\newblock {\it \bibinfo{journal}{SSRN Electronic Journal}\/},
  (p.~\bibinfo{pages}{33}). \DOIprefix\doi{10.2139/ssrn.2588523}.

\end{thebibliography}
